\documentclass[%
 reprint,
 amsmath,amssymb,
 aps,
pra,
]{revtex4-2}

\usepackage{graphicx}% Include figure files
\usepackage{dcolumn}% Align table columns on decimal point
\usepackage{bm}% bold math
\usepackage{hyperref}% add hypertext capabilities
\usepackage{epsfig}
\usepackage{color}
\usepackage{bbm}
\usepackage{longtable}

\def\ketm#1{  \left\vert  #1   \right\rangle   }

\def\mem#1#2#3{  \left\langle #1 \left\vert  #2 \right\vert #3 \right\rangle   }

\def\rmem#1#2#3{  \left\langle #1 \left\vert \left\vert  #2
                  \right\vert \right\vert #3 \right\rangle   }
\newcommand{\minus}[1]{ \mbox{$ (-1)^{#1} $}}

\newcommand{\CGC}[6]{ \mbox{$ \left( #1 #2 , #3 #4\,|\, #5 #6 \right) $} }

\newcommand{\SechsJ}[6]{ \mbox{$ %
            \arraycolsep0.25ex %
            \left\{ \begin{array}{ccc} %
                       #1 & #2 & #3 \vspace{0.5ex}\\%
                       #4 & #5 & #6 %
                   \end{array} \right\} $} }

\newcommand{\Dfun}[3] {\mbox {$D^{#1}_{#2}(#3)$}}

\begin{document}

\preprint{APS/123-QED}

\title{Angular distribution of photoelectrons generated in atomic ionization by twisted radiation}% Force line breaks with \\

\author{Maksim D. Kiselev}
\affiliation{ 
Skobeltsyn Institute of Nuclear Physics, Lomonosov Moscow State University, 119991 Moscow, Russia
}
 \affiliation{ 
 Faculty of Physics, Lomonosov Moscow State University, Moscow 119991, Russia
}
\affiliation{ 
 School of Physics and Engineering, ITMO University, Saint Petersburg, 197101 Russia
}
 \affiliation{ 
Laboratory for Modeling of Quantum Processes, Pacific National University, 680035 Khabarovsk, Russia
}%Lines break automatically or can be forced with \\
\author{Elena V. Gryzlova}
\affiliation{ 
Skobeltsyn Institute of Nuclear Physics, Lomonosov Moscow State University, 119991 Moscow, Russia
}
\author{Alexei N. Grum-Grzhimailo}%
 \affiliation{ 
Skobeltsyn Institute of Nuclear Physics, Lomonosov Moscow State University, 119991 Moscow, Russia
}
\affiliation{ 
 School of Physics and Engineering, ITMO University, Saint Petersburg, 197101 Russia
}

\date{\today}% It is always \today, today,

\begin{abstract}
Until recently, theoretical and experimental studies of photoelectron angular distributions (PADs) including non-dipole effects in atomic photo\-ionization have been performed mainly for the conventional plane-wave radiation. One can expect, however, that the non-dipole contributions to the angular- and polarization-resolved photo\-ionization properties are enhanced if an atomic target is exposed to twisted light. The purpose of the present study is to develop a theory for PADs to the case of twisted light, especially for many-electron atoms. The theoretical analysis is performed for the experimentally relevant case of macroscopic atomic targets, i.e., when the cross-sectional area of the target is larger than the characteristic transversal size of the twisted beam. For such a scenario, we derive expressions for the angular distribution of the emitted photoelectrons under the influence of twisted Bessel beams. As an illustrative example, we consider helium photo\-ionization in the region of the lowest dipole $2s2p\,[{^1P}_1]$ and quadrupole $2p^2\,[{^1D}_2]$ autoionization resonances. A noticeable variation of the PAD caused by changing the parameters of the twisted light is predicted.
\end{abstract}

%Use showkeys class option if keyword
\maketitle

%\tableofcontents

\section{Introduction}
\label{Intro}
Studies of photo-electron angular distributions (PADs) are important for many applications and are a fruitful source of  fundamental information about the structure of the target and its interaction with the radiation field. Until recently, theoretical analysis of the PADs was based, in accordance with most conventional experimental conditions, on the plane-wave approximation for the vector potential of the radiation field.
These studies were promoted to a new level when a twisted light became available in the vacuum ultraviolet (VUV) region \cite{Bahrdt2013}.
 In twisted radiation beams intensity, the profile has a nonuniform structure, since the surface of the constant phase differs from a plane and there are complex internal flow patterns~\cite{Mollina-Terriza2007,Bekshaev_2011}.
Recent developments in technology have made the generation of twisted radiation beams a routine procedure. They can be generated in different ways: with the use of spiral phase plates~\cite{Sueda:04, BEIJERSBERGEN1994321}, computer-generated holograms~\cite{Heckenberg:92}, $q$-plates~\cite{Karimi2009}, axicons~\cite{ARLT2000297}, integrated ring resonators~\cite{Cai2012}, or on-chip devices~\cite{Yang2021}. Twisted radiation beams can be generated in a broad range of energies from the optical region to the XUV range~\cite{Peele:02,Sasaki2008,Hemsing2011,Serbo2011,Serbo2011a,He:13, Bahrdt2013, Shen_2013,Ribic2014,hernandez2017}. The following types of twisted beams are mainly considered: Laguerre–Gaussian~\cite{ALLEN1992,ALLEN1999} and Bessel~\cite{Durnin:87,DurninPRL} beams. In the present work, we consider Bessel beams. Experimental (see the review~\cite{Babiker_2019}) and theoretical studies are being carried out on the interaction of twisted beams with matter: atoms~\cite{Surzhykov2016,Kosheleva2020,Ramakrishna2022}, molecules~\cite{Araoka2005,Peshkov2015}, and ions~\cite{MaH13,Seipt2016}.
Twisting of radiation brings new features into the interaction of the radiation with matter, which at present remain mostly unrevealed but are of great interest. For example, manifestation of non-dipole effects in photo\-ionization may strongly differ from those in the plane-wave case. Thus, a general theory for calculating PADs, especially for many-electron atoms, is required. In~\cite{MaH13}, a general formalism of ionic photo\-ionization by twisted Bessel light was developed with the use of hydrogen-like wave functions to describe the 
target system. It allowed to proceed very far without expanding the field in multipoles. In both cases, plane-wave and twisted beams, similar sets of field multipoles contribute to the PAD and, despite the fact that the selection rules are modified~\cite{Picon:10}, no radiation multipoles appear in comparison with plane waves. It is important to analyze the dependence of the observable quantities on the geometrical characteristics of the twisted beam and a target. It is known that in the angular distributions of photoelectrons in atomic photo\-ionization, the contribution of a certain interaction multipoles is associated with certain spherical harmonics~\cite{Cooper1990,Cooper1993}. In photo\-ionization by twisted beams, the PAD changes due to the redistribution of contributions from different multipoles. Thus, it is reasonable to assume that the PAD with twisted radiation exhibits the same structure as in the plane-wave approximation.  However, it differs (as will be shown below in present work) by some factors at the spherical harmonics, with these factors depending on the twisted light parameters. If this is the case, the PAD can serve as a diagnostic tool for twisted radiation beams. The photo\-excitation of atoms by Bessel beams was already discussed in~\cite{Schulz2019}.  The present work is in this sense its extension to the case of photo\-ionization.

This article is organized as follows. In Sec.~\ref{Section2} we review the general procedure for calculating the PAD in the case of photo\-ionization by plane-wave radiation. Section~\ref{Section3} is devoted to the development of the mathematical apparatus for calculating the PAD for photo\-ionization by twisted Bessel beams of different polarizations. In Sec.~\ref{Section4} we present an application of the developed approach to the photo\-ionization of helium in the vicinity of the lowest autoionizing resonances, together with a discussion of the PAD transformation near the spectroscopic features of the photo\-ionization cross section when changing the Bessel beam parameters. Unless stated otherwise, we use atomic units throughout.  
%%%%%%%%%%%%%%%%%%%%%%%%%%%%%%%%%%%%%%%%%%

\section{Plane-wave formalism}
\label{Section2}

Consider the atomic photo\-ionization process
\begin{equation}
\hbar\omega + A(\alpha_i J_i M_i) \rightarrow A^+(\alpha_f J_f M_f) + e^-({\bm p} m_s), \nonumber
\end{equation}
where $A$($A^+$) denotes the target atom before (after) ionization by a photon with energy $\hbar\omega$.

Below we show that \textit{any} theoretical analysis of the photo\-ionization of a many--electron atom by twisted light can be carried out using the matrix element
\begin{equation}
    \label{eq:matrix_element_plane_wave}
    M^{(\rm pl)}_{M_i \lambda M_f}\left({\bm k}, {\bm p} \right) = \mem{\alpha_f J_f M_f, \, {\bm p} m_s}{\hat{R}({\bm k}, \lambda)}{\alpha_i J_i M_i} \, ,
\end{equation}
which describes the photo\-ionization of an atom by a plane--wave photon with wave\-vector ${\bm k}$ and helicity $\lambda = \pm 1$. In the matrix element (\ref{eq:matrix_element_plane_wave}), $J_{i,f}$ and $M_{i,f}$ are the total angular momenta and their projections of the initial (before ionization) and final (after ionization) atomic (ionic) states, and $\alpha_{i,f}$ are all other quantum numbers needed for the state specification. The emitted electron is characterized by its momentum ${\bm p}$ and spin projections $m_s$ onto its propagation direction.

We first recall the well-known formalism for the description of the PAD in the plane wave photo\-ionization.

\subsection{Plane-wave matrix element}

\subsubsection{Electron-photon interaction operator}
$\hat{R}({\bm k}, \lambda)$ in the matrix element (\ref{eq:matrix_element_plane_wave}) is the inter\-action operator. In many-electron calculations, it can be written as a sum of single-particle operators:
\begin{equation}
    \label{eq:R_operator}
    \hat{R}({\bm k}, \lambda) = \sum\limits_{q} {\bm \alpha}_q {\bm u}_{\lambda} \, {\rm e}^{i {\bm k} {\bm r}_q} \, .
\end{equation}
Here ${\bm r}_q$ specifies the position of $q$--th electron, ${\bm u}_{\lambda}$ is the polarization vector and ${\bm \alpha}_q$ is the vector of the Dirac matrices for the $q^{\rm th}$ particle, and the summation is taken over all atomic electrons. Note that the operator (\ref{eq:R_operator}) is written in the relativistic framework for the Coulomb gauge for the electron--photon interaction. 

In order to evaluate the matrix element (\ref{eq:matrix_element_plane_wave}) with the operator (\ref{eq:R_operator}), it is practical to expand $\hat{R}({\bm k}, \lambda)$ into electric and magnetic multipole terms that are constructed as irreducible tensors of rank $L$. For the single-particle operator, this expansion is written as~\cite{Ros57}
\begin{eqnarray}
    \label{eq:multipole_expansion}
    {\bm \alpha} {\bm u}_{\lambda} \, {\rm e}^{i {\bm k} {\bm r}} &=& \sqrt{2\pi} \sum\limits_{LM} \sum\limits_{p=0,1} i^L \, \sqrt{2L + 1} \, \left(i \lambda \right)^{p} \times \nonumber\\
   &\times & D^L_{M \lambda}(\hat{{\bm k}}) \, {\bm \alpha} \, {\bm a}^{(p)}_{L M}({\bm r}) \, ,
\end{eqnarray}
where $\hat{{\bm k}} = (\theta_k, \phi_k)$ defines the direction of the incident (plane-wave) photon, $D^{j}_{m m'}(\hat{{\bm k}})$ is the Wigner D-function (see, for example,~\cite{Balashov2000}) and the magnetic ($p = 0$) and electric ($p = 1$) multipole terms are given by
\begin{eqnarray}
    \label{eq:multipole_fields}
 {\bm a}^{(p = 0)}_{L M}({\bm r}) &\equiv& {\bm a}^{(m)}_{L M}({\bm r}) = j_L(kr) \, {\bm T}^M_{L, L} \, ,
    \nonumber \\
 {\bm a}^{(p = 1)}_{L M}({\bm r}) &\equiv& {\bm a}^{(e)}_{L M}({\bm r}) = j_{L-1}(kr) \, \sqrt{\frac{L + 1 }{2L + 1}}{\bm T}^M_{L, L-1} - \nonumber\\
&&   - j_{L+1}(kr) \, \sqrt{\frac{L}{2L + 1}}{\bm T}^M_{L, L+1} \, .
\end{eqnarray}
Here, the vector spherical harmonics ${\bm T}^M_{L, \Lambda}$ are irreducible tensors of rank $L$, resulting from the coupling of the spherical unit vectors ${\bm e}_m$ ($m =0, \pm 1$) with the spherical harmonics:
\begin{eqnarray}
    \label{eq:vector_spherical_harmonics}
{\bm T}^M_{L, \Lambda}& =& \left[Y_{\Lambda} \otimes {\bm e} \right]_{LM} \equiv 
    \nonumber \\
  &  \equiv &\sum\limits_m \CGC{\Lambda}{M-m}{1}{m}{L}{M} Y_{\Lambda \, M-m}(\theta, \phi) \, {\bm e}_m \, .
\end{eqnarray}
%
%
%Based on this expression one can further evaluate the product:
%
%
%\begin{equation}
%    {\bm \alpha} \,{\bm a}^{(p)}_{L M}({\bm r}) = \left( \twobytwo{0}{{\bm \sigma}{\bm a}^{(p)}_{L M}({\bm r})}{{\bm \sigma}{\bm a}^{(p)}_{L M}({\bm r})}{0} \right) \, ,
%\end{equation}
%
%
%where ${\bm \sigma}{\bm a}^{(p)}_{L M}({\bm r})$ can be presented in terms of irreducible tensors $\left[Y_{\Lambda} \otimes {\bm \sigma} \right]_{LM}$.

\subsubsection{Continuum electron state}
Besides the electron--photon interaction operator, one needs to expand the wave-function of the emitted photo\-electron into partial waves:
\begin{equation}
    \label{eq:decomposition_continuum_electron}
    \ketm{{\bm p} m_s} = \sum\limits_{\kappa \mu} i^l \, {{\rm e}^{-i \Delta_\kappa}} \, [l] \, \CGC{l}{0}{\frac{1}{2}}{m_s}{j}{m_s} \, D^{j}_{\mu m_s}(\hat{\bm p}) \, \ketm{\epsilon \kappa \mu} \, ,
\end{equation}
where the summation runs over the Dirac angular-momentum quantum number $\kappa = \pm(j+1/2)$ for $l = j \pm 1/2$ with $l$ representing the orbital angular momentum of the electron waves $\ketm{\epsilon \kappa \mu}$. In Eq.~(\ref{eq:decomposition_continuum_electron}), we used standard notation for the Clebsch-Gordan coefficients, $\Delta_\kappa$ is the scattering phase, and the spin projection $m_s$ is defined with respect to the propagation direction $\hat{\bm p} = {\bm p}/p$. We also introduced the notation $[abc...] \equiv \sqrt{(2a+1)(2b+1)(2c+1)...}$\,. 

To obtain the partial-wave expansion of the \textit{many-electron} scattering states, we  use Eq.~(\ref{eq:decomposition_continuum_electron}) and apply the standard procedure for coupling two
angular momenta:

\begin{eqnarray}
 &&   \label{eq:many_electron_sattering_state}
    \ketm{\alpha_f J_f M_f, \, {\bm p} m_s} =\sum\limits_{\kappa \mu} i^l \, {{\rm e}^{-i \Delta_\kappa}} \, [l] \times \nonumber\\
  & \times& \CGC{l}{0}{\frac{1}{2}}{m_s}{j}{m_s} \, D^{j}_{\mu m_s}(\hat{\bm p}) \, \ketm{\epsilon \kappa \mu} \ketm{\alpha_f J_f M_f}= \nonumber \\
& = &\sum\limits_{\kappa \mu J_t M_t} i^l \, {{\rm e}^{-i \Delta_\kappa}} \, [l] \, \CGC{l}{0}{\frac{1}{2}}{m_s}{j}{m_s} \times \nonumber\\
& \times &\CGC{J_f}{M_f}{j}{\mu}{J_t}{M_t} \, D^{j}_{\mu m_s}(\hat{\bm p}) \,  \ketm{(\alpha_f J_f, \epsilon \kappa) J_t M_t} \, .
\end{eqnarray}
In expression (\ref{eq:many_electron_sattering_state}), the proper anti\-symmetrization of the outgoing electron with respect to all  bound-state orbitals should be ensured. 

\subsubsection{Evaluation of the plane-wave matrix element}
By using Eqs.~(\ref{eq:R_operator}) and (\ref{eq:many_electron_sattering_state}) and applying the Wigner-Eckart theorem, we obtain the matrix element for the plane-wave photons as

\begin{eqnarray}
    \label{eq:matrix_element_plane_wave_final}
 &&    M^{(\rm pl)}_{M_i \lambda M_f}\left({\bm k}, {\bm p} \right) = \sqrt{2\pi} \, \sum\limits_{LMp} \, \sum\limits_{\kappa \mu J_t M_t} \,  i^{-l + L} \, {{\rm e}^{i \Delta_\kappa}} \, \left(i \lambda \right)^{p} \nonumber\\
&&     \frac{[l][L]}{[J_t]} \, \CGC{l}{0}{\frac{1}{2}}{m_s}{j}{m_s} \,\CGC{J_f}{M_f}{j}{\mu}{J_t}{M_t}\times \nonumber\\
&&       \CGC{J_i}{M_i}{L}{M}{J_t}{M_t} 
     \, D^{j *}_{\mu m_s}(\hat{\bm p}) \, D^L_{M \lambda}(\hat{{\bm k}}) \nonumber\\
&&     \rmem{(\alpha_f J_f, \epsilon \kappa) J_t}{\sum\limits_{q} {\bm \alpha}_q \, {\bm a}^{(p)}_{L}({\bm r}_q)}{\alpha_i J_i} \, .
\end{eqnarray}

For the sake of brevity, we denote the many-body reduced matrix element as:
\begin{eqnarray}
    \label{eq:reduced_matrix_element_definition}
&&    \rmem{(\alpha_f J_f, \epsilon \kappa) J_t}{H_{\gamma}(pL)}{\alpha_i J_i} = \nonumber \\
& \!=\!&i^{-l} \, {{\rm e}^{i \Delta_\kappa}} \, \rmem{(\alpha_f J_f, \epsilon \kappa) J_t}{\sum\limits_{q} {\bm \alpha}_q \, {\bm a}^{(p)}_{L}({\bm r}_q)}{\alpha_i J_i}.
\end{eqnarray}
Using (\ref{eq:reduced_matrix_element_definition}), we  obtain

\begin{eqnarray}
 & &  \label{eq:matrix_element_plane_wave_final_2}
     M^{(\rm pl)}_{M_i \lambda M_f}\left({\bm k}, {\bm p} \right) = \sqrt{2\pi} \, \sum\limits_{LMp}  \, \sum\limits_{\kappa \mu J_t M_t} i^{L} \, \left(i \lambda \right)^{p} \nonumber\\
    && \frac{[lL] }{[J_t]} \,\CGC{l}{0}{\frac{1}{2}}{m_s}{j}{m_s} \,\CGC{J_f}{M_f}{j}{\mu}{J_t}{M_t}\times \nonumber\\
  &&    \CGC{J_i}{M_i}{L}{M}{J_t}{M_t}  \, D^{j *}_{\mu m_s}(\hat{\bm p}) \, D^L_{M \lambda}(\hat{{\bm k}}) \nonumber\\
     &&\rmem{(\alpha_f J_f, \epsilon \kappa) J_t}{H_{\gamma}(pL)}{\alpha_i J_i} \, .
\end{eqnarray}

\subsubsection{Photoelectron angular distribution}
We assume that the atom is initially unpolarized, and the polarization of
the residual ion and the photo\-electron spin are not detected. Therefore, we
average the PAD over the initial magnetic quantum numbers $M_i$ and sum over the
final magnetic quantum numbers $M_f$ and $m_s$:

\begin{equation} \label{eq:pad1}
W^{(\rm pl)}(\theta_p,\phi_p) = \frac{1}{2J_i+1} \sum_{M_i M_f m_s}
\left| M^{(\rm pl)}_{M_i \lambda M_f} \left({\bm k}, {\bm p} \right) \right|^2 \,.
\end{equation}

Defining the $z$-axis as the propagation direction of the incident photons,  
$\Dfun{L}{M \lambda}{\hat{{\bm k}}} = \Dfun{L}{M \lambda}{0,0,0} = \delta_{M \lambda}$.
Substituting this into Eqs.~(\ref{eq:matrix_element_plane_wave_final_2}) and (\ref{eq:pad1}),
we obtain

\begin{widetext}
\begin{eqnarray} \label{eq:10}
&&W(\theta_p, \phi_p)  =  \frac{2\pi}{2J_i+1} \sum_{M_i M_f m_s \atop{\lambda \lambda'}}
\sum_{Lp \atop{L'p'}}
\sum_{\kappa \mu \atop{\kappa' \mu'}} 
\sum_{J_t M_t \atop{J'_t M'_t}}
i^{L-L'} [LL'll'][J_t J'_t]^{-1} (i \lambda)^{p} (-i \lambda')^{p'}  
  \CGC{l}{0}{\frac{1}{2}}{m_s}{j}{m_s}  
 \CGC{l'}{0}{\frac{1}{2}}{m_s}{j'}{m_s}  \times \nonumber\\
&\times & \CGC{J_i}{M_i}{L}{\lambda}{J_t}{M_t} \, \CGC{J_i}{M_i}{L'}{\lambda'}{J'_t}{M'_t} 
   \CGC{J_f}{M_f}{j}{\mu}{J_t}{M_t} \, \CGC{J_f}{M_f}{j'}{\mu'}{J'_t}{M'_t} \,
\Dfun{j \ast}{\mu m_s}{\phi_p,\theta_p,0} \, \Dfun{j'}{\mu' m_s}{\phi_p,\theta_p,0} \times
\nonumber\\
&\times & \rmem{(\alpha_f J_f, \epsilon \kappa) J_t}{H_{\gamma}(pL)}{\alpha_i J_i} \,
\rmem{(\alpha_f J_f, \epsilon \kappa') J'_t}{H_{\gamma}(p'L')}{\alpha_i J_i}^{\ast} \,.
\end{eqnarray}
\end{widetext}

%\begin{widetext}
%\begin{eqnarray} \label{eq:11}
%\Dfun{j \ast}{\mu m_s}{\phi_p,\theta_p,0} \Dfun{j'}{\mu' m_s}{\phi_p,\theta_p,0}
%= \minus{m_s - \mu} \Dfun{j}{-\mu \, -m_s}{\phi_p,\theta_p,0} 
%\Dfun{j'}{\mu' m_s}{\phi_p,\theta_p,0} \nonumber\\
%= \minus{m_s - \mu} \sum_{kq} \CGC{j'}{\mu'}{j}{-\!\mu}{k}{q} 
%\CGC{j'}{m_s}{j}{-\!m_s}{k}{0} \Dfun{k}{q0}{\phi_p,\theta_p,0}
%\end{eqnarray}
%\end{widetext}

%
%
%
%By using this expansion we can perform summation over $\mu', \mu, M_i$
%\begin{eqnarray} \label{eq:12}
%\sum_{\mu \mu' M_i} \minus{m_s-\mu} \CGC{j'}{\mu'}{j}{-\mu}{k}{q} 
%\CGC{J_i}{M_i}{L}{\lambda}{J_t}{M_t} 
%\CGC{J_i}{M_i}{L'}{\lambda}{J'_t}{M'_t} =...= \nonumber\\
%= \minus{J_t-J'_t+k+J_i+m_s+1} [J_t J'_t] 
%\CGC{L}{\lambda}{L'}{-\lambda}{k}{0} 
%\SechsJ{J'_t}{J_t}{k}{L}{L'}{J_i} \delta_{q0} \quad phase!
%\end{eqnarray}
%
%We can now perform also summation over $m_s$
%\begin{eqnarray} \label{eq:13}
%\sum_{m_s} \minus{J_i+m_s} \CGC{j'}{m_s}{j}{-m_s}{k}{0} 
%\CGC{l}{0}{\frac{1}{2}}{m_s}{j}{m_s} 
%\CGC{l'}{0}{\frac{1}{2}}{m_s}{j'}{m_s} = ... =
%\nonumber\\
%= \minus{J_i-\frac{1}{2}}[jj'] \CGC{l}{0}{l'}{0}{k}{0} 
%\SechsJ{j'}{j}{k}{l}{l'}{\frac{1}{2}} \quad phase!
%\end{eqnarray}
%
We can further proceed as follows: evaluate the product of two Wigner $D$-functions; sum over $M_i M_f M_t M'_t$
by using the first equality given in~(A.91) of~\cite{Balashov2000};
sum over $\mu \mu'$ using the unitarity of the Clebsch-Gordan coefficients; and sum over $m_s$ using Eq.~(A.90) of~\cite{Balashov2000}. After these steps, we finally obtain

%By combining (\ref{eq:12}), (\ref{eq:13}) together with Eq.~(\ref{eq:10}) we find
\begin{widetext}
\begin{eqnarray} \label{eq:14}
W(\theta_p, \phi_p)  =  
\frac{2\pi}{2J_i+1} \sum_{kq} \Dfun{k}{q0}{\phi_p,\theta_p,0} \sum_{LpL'p' \atop{J_t J'_t \kappa \kappa'}} \sum_{\lambda \lambda'} i^{L-L'} [LL'll'jj'J_tJ'_t] (i \lambda)^{p} (-i \lambda')^{p'} 
  \minus{L+\lambda+J_i-J_f+\frac{1}{2}} \, 
 \CGC{L}{\lambda}{L'}{-\!\lambda'}{k}{q} \times \nonumber \\
 \times  \CGC{l}{0}{l'}{0}{k}{0} \,
  \SechsJ{j}{j'}{k}{l'}{l}{\frac{1}{2}} \SechsJ{J_t}{J'_t}{k}{L'}{L}{J_i}
  \SechsJ{J_t}{J'_t}{k}{j'}{j}{J_f} \rmem{(\alpha_f J_f, \epsilon \kappa) J_t}{H_{\gamma}(pL)}{\alpha_i J_i} \,
\rmem{(\alpha_f J_f, \epsilon \kappa') J'_t}{H_{\gamma}(p'L')}{\alpha_i J_i}^{\ast} \, .
\end{eqnarray}
\end{widetext}

\section{Twisted-wave formalism}
\label{Section3}
This section is devoted to the discussion of photo\-ionization by  twisted light.
\subsection{Evaluation of the twisted-wave matrix element}
\begin{figure}
\centering
\includegraphics[width=0.49\textwidth]{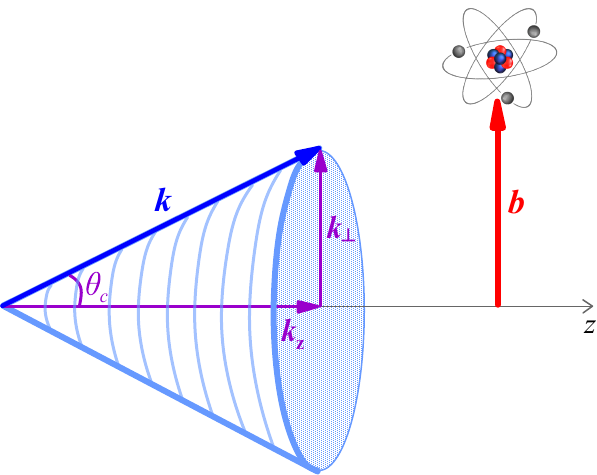}
\caption{The overview of the Bessel beam parameters and position of a target atom.\label{fig0}}
\end{figure}

We assume that the light is prepared in a so-called Bessel state. In our analysis, the Bessel photon beam propagates along the (quantization) $z$~axis. For this case, the Bessel state is characterized by the well-defined projections of the linear momentum $k_z$ and the total angular momentum (TAM) onto the $z$~axis $m_{\mathsf{tam}}$. The absolute value of the transverse momentum, $\kappa_{\perp} = \left| {\bf k}_{\perp} \right|$, is fixed; together with $k_z$ it defines the energy of the photons $\omega = c \sqrt{\kappa_{\perp}^2 + k_z^2}$. As shown in~\cite{MaH13}, this Bessel state is described by the vector potential 
\begin{equation} 
    \label{eq:18}
    {\bf A}_{\kappa_{\perp} m_{\mathsf{tam}} \lambda}^{tw} = \int {\bf u}_{\lambda} e^{i{\bf kr}} 
    a_{\kappa_{\perp} m_{\mathsf{tam}}}({\bf k}_{\perp}) \frac{d^2 {\bf k}_{\perp}}{4 \pi^2} \, ,
\end{equation}
where
\begin{equation} 
    \label{eq:18a}
    a_{\kappa_{\perp} m_{\mathsf{tam}}}({\bf k}_{\perp}) =  (-i)^{m_{\mathsf{tam}}} e^{i m_{\mathsf{tam}} \phi_k} \sqrt{\frac{2 \pi}{k_{\perp}}}
    \delta(k_{\perp} - \kappa_{\perp}) \, .
\end{equation}
 These expressions present the Bessel state in momentum space as a coherent superposition of plane waves with their wave vectors ${\bm k} = \left({\bm k}_\perp, k_z\right)$ lying on the surface of a cone with opening angle $\tan\theta_c = k_\perp/k_z$ (see Fig.~\ref{fig0}). Below we characterize the kinematic properties of the Bessel beams by this opening angle.

Using the vector potential (\ref{eq:18}), (\ref{eq:18a}), we can derive the matrix element for photo\-ionization of a many-electron atom by twisted light:  
\begin{eqnarray}
    \label{eq:matrix_element_twisted}
    && M^{(\rm tw)}_{M_i \lambda \, m_{\mathsf{tam}} M_f}\left({\bm p}; \theta_c, \, {\bm b} \right) = \nonumber\\
  &  =&\int \, a_{\kappa_{\perp} m_{\mathsf{tam}}}({\bf k}_{\perp}) \, {\rm e}^{-i {\bm k}_\perp {\bm b}} \, M^{(\rm pl)}_{M_i \lambda M_f}\left({\bm k}, {\bm p} \right)\frac{d^2 {\bf k}_{\perp}}{4 \pi^2}  \,  \, ,
\end{eqnarray}
where $M^{(\rm pl)}_{M_i \lambda M_f}\left({\bm k}, {\bm p} \right)$ is the conventional plane-wave matrix element (\ref{eq:matrix_element_plane_wave}). We introduced an additional exponential factor ${\rm e}^{-i {\bm k}_\perp {\bm b}}$ to specify the lateral position of the target atom with regard to the beam axis of the incident light, where the impact parameter ${\bm b} = \left(b_x, b_y, b_z \right)$. This parameter is essential since, in contrast to a plane wave, the Bessel beams have a much more complex internal structure. In particular, their intensity distribution in the transverse direction (the $xy$~plane) is not uniform but consists of concentric rings of high and low intensity. The direction of the local energy flux also varies significantly within the wavefront. Therefore, one may  expect that the properties of the photo\-electrons strongly depend on the position $\bm b$ of the target atom with respect to the Bessel beam axis.

\subsection{Differential cross section for ionization by twisted light}

With the help of the ``twisted'' matrix element (\ref{eq:matrix_element_twisted}), one can evaluate the 
angle-differential photo\-ionization cross section. This cross section depends on both the polarization state of the incident photons and the spatial arrangement of the target. We start our analysis from the simplest case of a \textit{homogeneous macroscopic} target that consists of atoms randomly and uniformly distributed within the $xy$~plane. Below we consider the differential cross section for various polarization states of the incident light for such a target.

\subsubsection{Circularly polarized light}
To evaluate the angle-differential cross section for ionization of the macroscopic atomic target by twisted light, we have to average the squared matrix element (\ref{eq:matrix_element_twisted}) over the impact parameter:
\begin{eqnarray}
    \label{eq:crpss_section_circular}
    &&\frac{{\rm d}\sigma^{\rm (tw, circ)}}{{\rm d}\Omega_p}\left(\theta_p, \phi_p ; \, \theta_c\right) = \mathcal{N} \frac{1}{2J_i + 1} \times \nonumber\\
  &\times & \sum\limits_{M_i M_f m_s}
    \int \left| M^{(\rm tw)}_{M_i \lambda \, m_{\mathsf{tam}} M_f}\left({\bm p}; \theta_c, \, {\bm b} \right) \right|^2 \, \frac{{\rm d}{\bm b}}{\pi R^2} \, .
\end{eqnarray}
Here $R$ defines the ``size'' of the target, which is assumed to be much larger than the characteristic size of the intensity rings in the Bessel beam. We assume that the atom in the initial state is unpolarized and the polarization state of neither the ion nor the electron spin are  detected. The evaluation of the prefactor $\mathcal{N}$ is not a trivial task since it requires the re\-definition of the concept of a cross section for the case of twisted light. Here we will follow the concept of the cross section defined in Ref.~\cite{ScF14}.

By inserting the matrix element (\ref{eq:matrix_element_twisted}) into Eq.~(\ref{eq:crpss_section_circular}) and carrying out the necessary algebra (see Appendix~\ref{AppendixA} for details), we obtain the expression for the differential cross-section in the case of circularly polarized Bessel light:
\begin{widetext}
\begin{eqnarray} \label{eq:cr11}
 \frac{{\rm d}\sigma^{\rm (tw, circ)}}{{\rm d}\Omega_p}\left(\theta_p, \phi_p ; \, \theta_c\right)  &  = &  \frac{2 \pi \mathcal{N}}{2J_i + 1} 
 \sum_J \minus{J} P_J(\cos \theta_p) P_J(\cos \theta_c)
  \sum_{\kappa \kappa' J_t J'_t} \sum_{LL'pp'} i^{L-L'} [ll'J_tJ'_t] [LL']^2 \times \nonumber \\
& \times & (i \lambda)^{p-p'} 
\minus{\lambda+J_i-J_f+\frac{1}{2}} \, 
 \CGC{L}{\lambda}{L'}{-\!\lambda}{J}{0} \, \CGC{l}{0}{l'}{0}{J}{0} 
 %\nonumber \\ 
  \SechsJ{j}{j'}{J}{l'}{l}{\frac{1}{2}} \SechsJ{J_t}{J'_t}{J}{L'}{L}{J_i}
  \SechsJ{J_t}{J'_t}{J}{j'}{j}{J_f} \times \nonumber\\  
 & \times & \rmem{(\alpha_f J_f, \epsilon \kappa) J_t}{H_{\gamma}(pL)}{\alpha_i J_i} \,
\rmem{(\alpha_f J_f, \epsilon \kappa') J'_t}{H_{\gamma}(p'L')}{\alpha_i J_i}^{\ast} .
\end{eqnarray}
\end{widetext}

The limiting case of plane waves may be obtained from (\ref{eq:cr11}) by putting $\theta_c=0$. Then $P_J(\cos \theta_c) = 1$ and comparing  with (\ref{eq:14}), we see that these equations coincide. Thus, we obtain a result, which can be formulated as

\noindent
\textbf{Statement 1}:
\textit{For circularly polarized Bessel beams and an extended atomic target,
the effect of twisting on the photoelectron angular distribution is expressed by multiplying each coefficient
of the Legendre polynomial $P_k(\cos \theta_p)$ by a factor $P_k(\cos \theta_c)$, where $\theta_c$
is the opening angle of the twisted radiation cone. This result is independent of the field multipoles and the target structure.}

\subsubsection{Linearly polarized light}
In the previous subsection, we considered the ionization of an atom by a twisted light, characterized by the well--defined values of the TAM projection $m_{\mathsf{tam}}$ and the helicity $\lambda$. In the paraxial limit, which corresponds to small values of the opening angle $\theta_c \ll 1$, that case corresponds to the well-known case of \textit{circularly} polarized light. Now we turn to the case that can be considered as a twisted \textit{analogon} of plane-wave linearly polarized light. The vector potential of twisted light  that is ``linearly polarized'' in $xz$ plane can be written as
\begin{eqnarray}
    \label{eq:twisted_linearly_polarized}
    {\bf A}_{\kappa_{\perp} \, \parallel}^{tw} = \frac{i}{\sqrt{2}} \, ( 
    {\bf A}_{\kappa_{\perp} m_{\mathsf{tam}} = m_{\mathsf{oam}} + 1, \, \lambda = +1}^{tw} - \nonumber\\
    - {\bf A}_{\kappa_{\perp} m_{\mathsf{tam}} = m_{\mathsf{oam}} - 1, \, \lambda = -1}^{tw} ) \, ,
\end{eqnarray}
i.e., as the difference of two vector potentials (\ref{eq:18}), obtained for different TAM projections and different helicities. The physical meaning of Eq.~(\ref{eq:twisted_linearly_polarized}) becomes transparent if one writes this expression in the paraxial regime
\begin{eqnarray}
    \label{eq:twisted_linearly_polarized_paraxial}
    {\bf A}_{\kappa_{\perp} \, \parallel}^{tw} &\approx& {\bm e}_x \, J_{m_{\mathsf{oam}}}(\kappa_\perp r_\perp) \, {\rm e}^{i m_{\mathsf{oam}} \phi} \, 
    {\rm e}^{i k_z z} \, ,
\end{eqnarray}
where we applied the approach from Ref.~\cite{MaH13}.  Here, $m_{\mathsf{oam}}$ can be considered as the projection of the light's orbital angular momentum (OAM). 

Using the general expression for the vector potential of linearly polarized radiation (\ref{eq:twisted_linearly_polarized}) allows us to derive the photo\-ionization matrix element
\begin{eqnarray}
    \label{eq:matrix_element_photoionization_linear}
 & &    M^{(\rm tw)}_{M_i M_f \, \parallel}\left({\bm p}; \theta_c, \, {\bm b} \right)=\nonumber\\
   &=&  \frac{i}{\sqrt{2}} \, \Big( M^{(\rm tw)}_{M_i \lambda = + 1\, m_{\mathsf{tam}} = m_{\mathsf{oam}} +1 \, M_f}\left({\bm p}; \theta_c, \, {\bm b} \right) \nonumber\\
   &  -& M^{(\rm tw)}_{M_i \lambda = - 1\, m_{\mathsf{tam}} = m_{\mathsf{oam}} - 1 \, M_f}\left({\bm p}; \theta_c, \, {\bm b} \right)\Big) \, ,
\end{eqnarray}
in terms of the matrix elements (\ref{eq:matrix_element_twisted}). Applying this expression and performing algebra similar to that in the previous subsection, we can derive the differential cross section for the ionization by ``linearly polarized'' twisted light:
\begin{widetext}
\begin{eqnarray}
    \label{eq:crpss_section_linear}
    \frac{{\rm d}\sigma^{\rm (tw, lin)}}{{\rm d}\Omega_p}\left(\theta_p, \phi_p ; \, \theta_c\right) &=& \mathcal{N} \frac{1}{2J_i + 1} \, \sum\limits_{M_i M_f m_s}
    \int \left| M^{(\rm tw)}_{M_i M_f \, \parallel}\left({\bm p}; \theta_c, \, {\bm b} \right)  \right|^2 \, \frac{{\rm d}{\bm b}}{\pi R^2} = \nonumber \\
 &   =& \mathcal{N} \frac{1}{2J_i + 1}  \, \sum\limits_{M_i M_f m_s} 
    \sum\limits_{\lambda, \lambda'} \int M^{(\rm pl)}_{M_i \, \lambda \, M_f}({\bm k}, {\bm p}) \, M^{(\rm pl) *}_{M_i \, \lambda' \, M_f}({\bm k}, {\bm p}) \, {\rm e}^{i (\lambda - \lambda') \varphi_k}  \, \frac{d {\varphi_k}}{2 \pi}  \, .
\end{eqnarray}
\end{widetext}
For $\theta_c \to 0$, this expression reduces to the well-known plane-wave result. Indeed, by using the asymptotic expression for the Wigner D-function
\begin{equation}
    D^L_{M \lambda}(\varphi_k, \theta_c, 0) \approx {\rm e}^{-i \lambda \varphi_k} \, \delta_{\lambda M} \, , 
\end{equation}
we can write the plane-wave photo\-ionization matrix element as

\begin{eqnarray}
    \label{eq:M_asymptotic}
    M^{(\rm pl)}_{M_i \, \lambda \, M_f}({\bm k}, {\bm p}) \equiv 
    M^{(\rm pl)}_{M_i \, \lambda \, M_f}(\theta_c, \varphi_k , {\bm p}) \approx \nonumber\\
   \approx {\rm e}^{-i \lambda \varphi_k}  M^{(\rm pl)}_{M_i \, \lambda \, M_f}(\theta_c = 0, \varphi_k = 0, {\bm p}) \equiv \nonumber\\
  \equiv {\rm e}^{-i \lambda \varphi_k} \, M^{(\rm pl)}_{M_i \, \lambda \, M_f}(0, {\bm p}) \, ,
\end{eqnarray}
where the last matrix element describes plane-wave radiation propagating along quantization $z$ axis. Substituting Eq.~(\ref{eq:M_asymptotic}) into Eq.~(\ref{eq:crpss_section_linear}) we obtain

\begin{eqnarray}
  &&  \label{eq:crpss_section_linear_plane_wave}
    \frac{{\rm d}\sigma^{\rm (pl, lin)}}{{\rm d}\Omega_p}\left(\theta_p, \phi_p ; \, \theta_c\right) = \mathcal{N} \frac{1}{2J_i + 1}  \times \nonumber\\
    &\times & \sum\limits_{M_i M_f m_s} 
    \sum\limits_{\lambda, \lambda'} M^{(\rm pl)}_{M_i \, \lambda \, M_f}(0, {\bm p}) \, M^{(\rm pl) *}_{M_i \, \lambda' \, M_f}(0, {\bm p})  \, .
\end{eqnarray}

%
%
%We stress that this expression corresponds to the relation between linear and circular polarization vectors in Rose convention, see~\cite{Ros57}:
%
%
%\begin{eqnarray}
%    {\bm u}_x &=& \frac{1}{\sqrt{2}} \, \left( {\bm u}_{+1} + {\bm u}_{-1} \right) \, , \\[0.2cm]
%    {\bm u}_y &=& \frac{i}{\sqrt{2}} \, \left( {\bm u}_{-1} - {\bm u}_{+1} \right) \, .
%\end{eqnarray}

To simplify Eq.~(\ref{eq:crpss_section_linear}), we first substitute the matrix elements (\ref{eq:matrix_element_plane_wave_final_2}) and after further transformations (see Appendix~\ref{AppendixB} for details), we obtain the expression for the differential cross-section in the case of linearly polarized Bessel light:

\begin{widetext}
\begin{eqnarray}
\label{eq:PADlin_fin}
 \frac{{\rm d}\sigma^{\rm (tw, lin)}}{{\rm d}\Omega_p}\left(\theta_p, \phi_p ; \, \theta_c\right)& = &\frac{2 \pi \mathcal{N}}{2J_i + 1} 
\sum_J \sum_{M=0, \pm2} \sqrt{\frac{4\pi}{2J+1}} Y^{\ast}_{JM}(\theta_p, \phi_p) \, d^J_{MM}(\theta_c) \sum_{\lambda, \lambda' } 
\delta_{M, \lambda-\lambda'} \times \nonumber\\
&\times & \sum_{\kappa \kappa' J_t J'_t}
\sum_{LL'pp'} i^{L-L'} (i \lambda)^p (-i \lambda')^{p'} \minus{J_i-J_f+\frac{1}{2}} \minus{L+\lambda} [ll'LL'J_t J'_t] \CGC{l}{0}{l'}{0}{J}{0} \CGC{L}{\lambda}{L'}{-\!\!\lambda'}{J}{M} \times \nonumber \\
&\times & \SechsJ{j}{j'}{J}{l'}{l}{\frac{1}{2}}
\SechsJ{J_t}{J'_t}{J}{L'}{L}{J_i} \SechsJ{J_t}{J'_t}{J}{j'}{j}{J_f}
 \rmem{(\alpha_f J_f, \epsilon \kappa) J_t}{H_{\gamma}(pL)}{\alpha_i J_i} \,
\rmem{(\alpha_f J_f, \epsilon \kappa') J'_t}{H_{\gamma}(p'L')}{\alpha_i J_i}^{\ast} .
\end{eqnarray}
\end{widetext}

In Eq.~(\ref{eq:PADlin_fin}) $d^j_{mm'}(\theta)$ is the small Wigner D-function,~\cite{Balashov2000}. For $\theta_c=0$, $d^J_{MM}(0) = 1$, and thus we can make a second statement.
\\
\noindent\textbf{Statement 2}:
\textit{For linearly polarized Bessel beams and an extended atomic target, the effect of twisting on the photoelectron angular distribution is expressed by multiplying each coefficient of the spherical harmonic $Y_{kq}(\theta_p,\phi_p)$ by a factor  $d^k_{qq}(\theta_c)$, where $\theta_c$ is the opening angle of the twisted radiation cone. The result is independent of the field multipoles and the atomic structure.}

%%%%%%%%%%%%%%%%%%%%%%%%%%%%%%%%%%%%%%%%%%
\section{Results and discussion}
\label{Section4}
The main consequence of \textbf{Statements 1} and \textbf{2} is the possibility to use the well-known parameterization of the PADs in photo\-ionization by plane-wave radiation in terms of the anisotropy parameters $\beta, \gamma$, and $\delta$~\cite{Cooper1990} also for the case of photo\-ionization by twisted radiation. The difference between ``plane" and ``twisted" PADs comes down to geometrical multipliers depending on the angle of the twisted radiation cone $\theta_c$, while the anisotropy parameters remain unaffected.  This makes it possible to calculate them by different methods and models. It also means that one may expect a more pronounced change in the PAD when at least one of the anisotropy parameters changes significantly.

\begin{figure}
\centering
\includegraphics[width=0.49\textwidth]{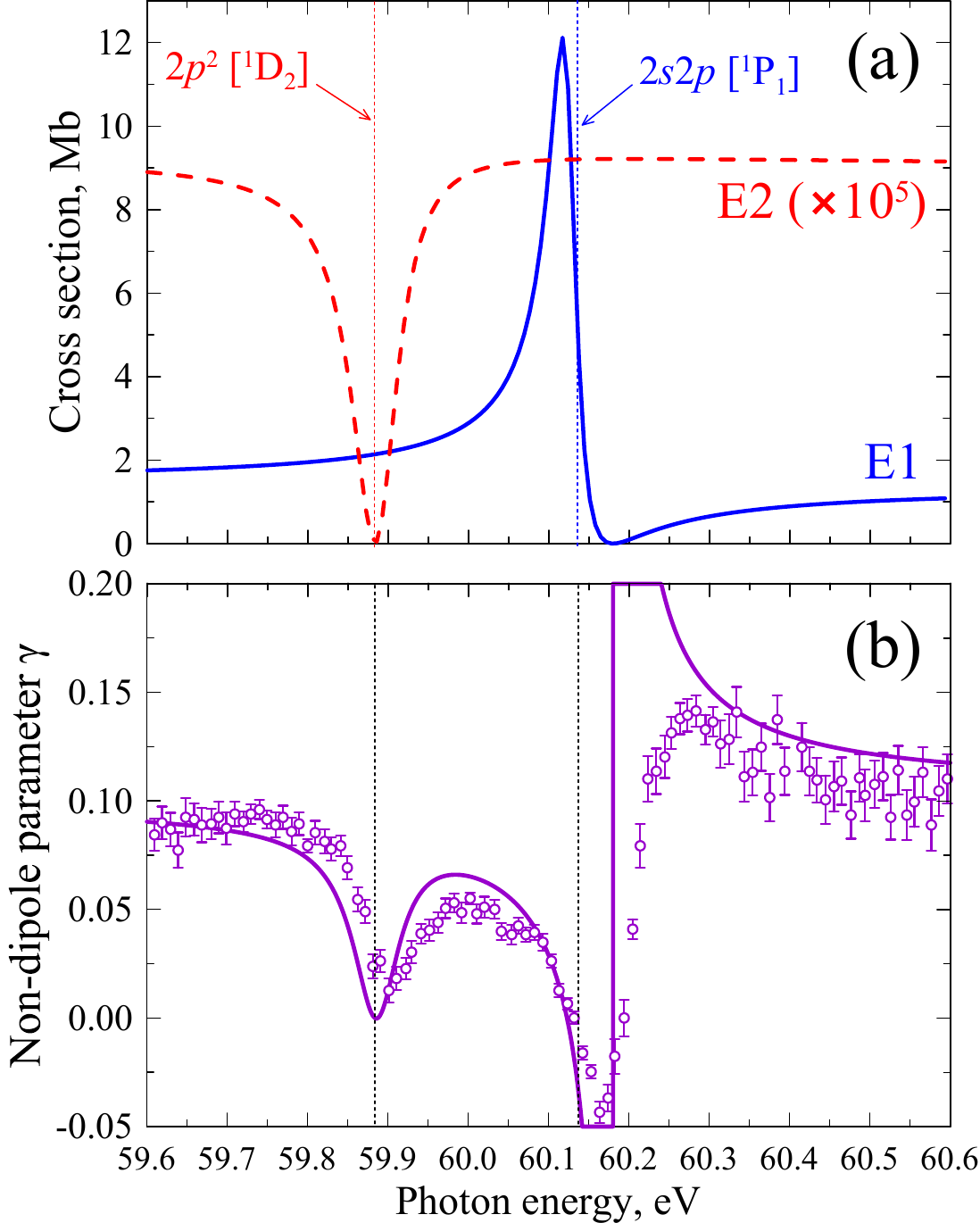}
\caption{(\textbf{a}) Cross sections for dipole $E1$ (solid blue line) and quadrupole $E2$ (red dashed line) photo\-ionization of helium. (\textbf{b}) Non-dipole parameter $\gamma$ of the photoelectron angular distribution.  The solid line represents the theoretical prediction, the open circles are the experimental data from~\cite{Krassig2002}.\label{fig1}}
\end{figure}

For an illustration of our ideas, we now consider photo\-ionization of the helium atom in the vicinity of the lowest autoionization states (AIS): dipole $2s2p\,[{^1P}_1]$ and quadrupole $2p^2\,[{^1D}_2]$, respectively. The calculation of the dipole and quadrupole photo\-ionization amplitudes was performed by means of the B-spline R-matrix code \cite{Zatsarinny2006} within the $LS$-coupling scheme to  describe the initial atomic and final ionic states. All the wave functions were obtained by the multi\-configuration Hartree-Fock method using the MCHF 
code~\cite{Froese97}. For the initial state, we first performed a Hartree-Fock (HF) optimization of the $1s^2\,[{^1S}]$ state in order to obtain a first approximation of the $1s$-orbital. Then we added the configurations of the same parity $1s2s, 1s3s, 2s^2, 2s3s, 2p^2, 2p3p, 3p^2, 3d^2$ to the ground-state description, optimizing all of them together on the $^1S$ term. For the six final ionic states (targets) we used single-configuration representations $1s\,[{^2S}], 2s\,[{^2S}], 2p\,[{^2P}], 3s\,[{^2S}], 3p\,[{^2P}], 3d\,[{^2D}]$. The dipole and quadrupole photo\-ionization cross sections are presented in Fig.~\ref{fig1}a.

It is well-known that for ionization of an $s$-shell $\beta=2$ and $\delta=0$.  Therefore, $\gamma$ remains the only parameter that may change with the photon energy. The non-dipole parameter $\gamma$ characterizes the interference between the electric dipole ($E1$) and quadrupole ($E2$) photo\-ionization amplitudes. One should expect the sharpest modulation of this parameter when the $E2$ photo\-ionization cross section becomes comparable to or even dominates  $E1$ photo\-ionization. For example, such a situation is observed in helium photo\-ionization near dipole the $2s2p\,[{^1P}_1]$ and quadrupole $2p^2\,[{^1D}_2]$ AIS resonances. At the photon energy of $\approx 60.18$~eV, the cross section of the $E1$ dipole photo\-ionization approaches zero, i.e., $\sim 10^{-4}$~Mb.  Hence the $E2$ quadrupole photo\-ionization dominates in this region (see Fig.~\ref{fig1}a).

The photon energy-dependence of $\gamma$ is presented in Fig.~\ref{fig1}b together with experimental data points from \cite{Krassig2002}. Comparison of the present theoretical results with experimental data shows  a significant discrepancy around $60.2$~eV. This issue was extensively studied 
in~\cite{Argenti2010}.  The most probable and plausible reason for such a difference is an underestimation of the background signal, i.e., an entirely instrumental origin during the experimental data processing. The spectroscopic models, however, appear to be reliable.

%\unskip

\begin{figure}
\centering
\includegraphics[width=0.49\textwidth]{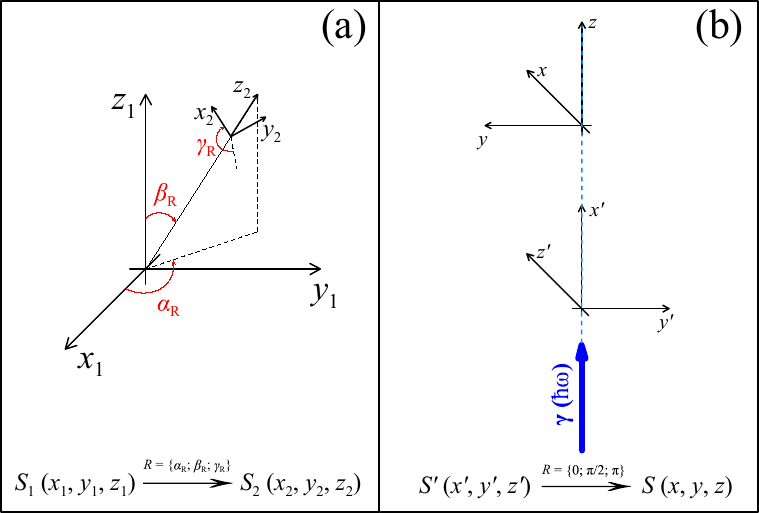}
\caption{(\textbf{a}) General transformation between coordinate systems $S_1$ and $S_2$ by the Euler angle rotation $R=\{\alpha_R; \beta_R; \gamma_R\}$.
(\textbf{b}) The transformation between the coordinate systems $S'$ (used in  Eq.~(\ref{eq:PADplane_generalS'})) and $S$ (used in the Eq.~(\ref{eq:PADplane_generalS})) is provided by $R=\{0;\pi/2;\pi\}$. %In this panel $\gamma (\hbar\omega)$ denotes either plane or twisted wave considering blue thick arrow to be aligned with the wave--vector ${\bm k}$ or with the twisted cone axis, respectively.
\label{fig2_3}}
\end{figure}

In order to evaluate the expression for the PAD resulting from twisted wave photo\-ionization, we start from the well-known plane-wave PAD in the general form within first-order non-dipole corrections~\cite{Shaw1996}. We choose the coordinate system $S'$ (see Fig.~\ref{fig2_3}) in such a way that the $x'$-axis is the propagation axis of the light beam ($\bm{k}\parallel x'$) and the $z'$-axis is the polarization axis:

\begin{widetext}
\begin{eqnarray} \label{eq:PADplane_generalS'}
\left( \frac{d\sigma}{d\Omega} \right)_{S'} = \frac{\sigma_0}{4\pi} \left( \left[1+\frac{\beta}{4}-\frac{3}{4}P\beta+\frac{3}{2}P\beta\cos^2\theta'_p \right] + \left[\delta+\gamma P\cos^2\theta'_p-\gamma\frac{P-1}{2} \right]\sin\theta'_p\cos\phi'_p+\right. \nonumber\\
& & \hspace*{-10.5cm} \left.+\frac{3\beta}{4}(P-1)(\sin\theta'_p\cos\phi'_p)^2 + \gamma\frac{P-1}{2}(\sin\theta'_p\cos\phi'_p)^3 
\right),
\end{eqnarray}
\end{widetext}
Here $P$ is the degree of linear polarization. $P=1$ in Eq.~(\ref{eq:PADplane_generalS'}) corresponds to the case of linearly polarized light, while $P=0$ corresponds to the case of circularly polarized light. For  convenience, one can express all the combinations of sines and cosines in Eq.~(\ref{eq:PADplane_generalS'}) in terms of spherical harmonics $Y_{lm}(\theta'_p,\phi'_p)$. To apply the statements obtained in Section~\ref{Section3}, we need to transform Eq.~(\ref{eq:PADplane_generalS'}) from the coordinate system $S'$ to $S$, where the $x$-axis is the polarization axis and the $z$-axis is the propagation direction. The conversion of spherical harmonics when coordinate system undergoes rotation described by the triad of Euler angles  $R=\{\alpha_R;\beta_R;\gamma_R\}$ is given by~\cite{Balashov2000}
\begin{equation}
\label{eq:harmonics_transformation}
Y_{KQ'}(\theta'_p,\phi'_p)=\sum_{Q}D^K_{QQ'}(\alpha_R;\beta_R;\gamma_R)Y_{KQ}(\theta_p,\phi_p).
\end{equation}
The transformation $S'\rightarrow S\,(z'\rightarrow x;\,x'\rightarrow z)$ is provided by the rotation $R=\{0;\frac{\pi}{2};\pi\}$. The  transformation of Eq.~(\ref{eq:PADplane_generalS'}) then leads to

\begin{widetext}
\begin{eqnarray} \label{eq:PADplane_generalS}
\left( \frac{d\sigma}{d\Omega} \right)_{S} = \frac{\sigma_0}{4\pi} \left( 1-\frac{\beta}{2}\sqrt{\frac{4\pi}{5}}\left[Y_{20}(\theta_p,\phi_p)-\frac{P\sqrt{6}}{2}\left(Y_{2-2}(\theta_p,\phi_p)+Y_{2+2}(\theta_p,\phi_p) \right) \right] + \left(\delta+\frac{\gamma}{5} \right)\sqrt{\frac{4\pi}{3}}Y_{10}(\theta_p,\phi_p)- \right. \nonumber\\
& & \hspace*{-10.5cm} \left. - \frac{\gamma}{5}\sqrt{\frac{4\pi}{7}}\left[Y_{30}(\theta_p,\phi_p)-P\sqrt{\frac{5}{6}}\left(Y_{3-2}(\theta_p,\phi_p)+Y_{3+2}(\theta_p,\phi_p) \right) \right] 
\right).
\end{eqnarray}
\end{widetext}

\begin{figure*}
\begin{center}
\includegraphics[width=17.5 cm]{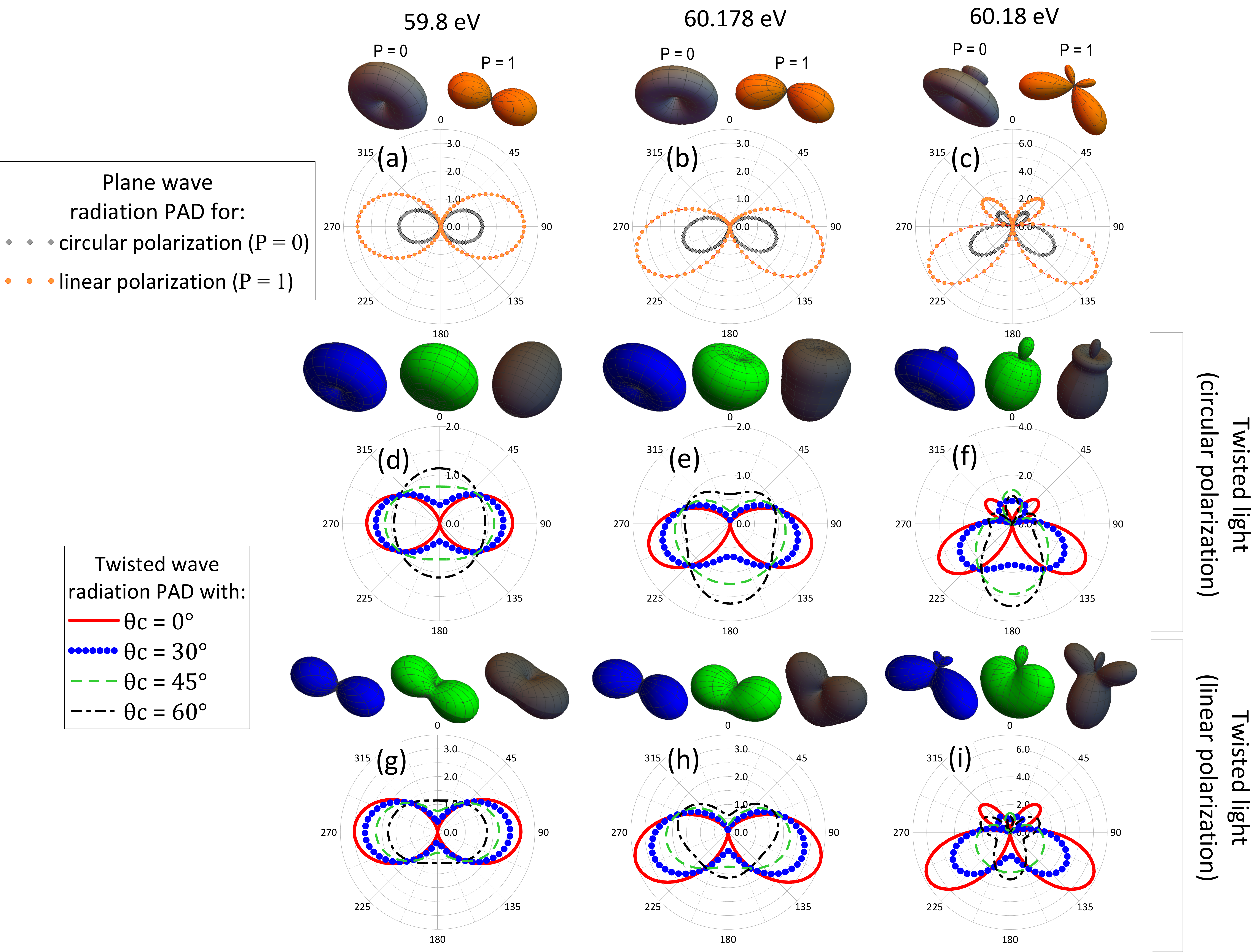}
\caption{(\textbf{a,b,c}) Simulated, according to Eq.~(\ref{eq:PADplane_generalS}), PADs for plane-wave photo\-ionization of helium (diamonds and circles along the line indicate the angular grid step). The upper part shows 3D views of the angular distributions for circularly polarized ($P=0$, gray one on the left) and linearly polarized ($P=1$, orange one on the right) light. (\textbf{d,e,f}) Simulated, according to \textbf{Statement 1}, PADs for twisted circularly polarized Bessel beam photo\-ionization of helium for different values of~$\theta_c$. The upper part shows 3D views of the angular distributions for $\theta_c=30^{\circ};\,45^{\circ}$ and $60^{\circ}$ (from left to right). (\textbf{g,h,i}) Simulated, according to \textbf{Statement 2}, PADs for twisted linearly polarized Bessel beam photo\-ionization of helium for different values of~$\theta_c$. The upper part shows 3D views of the angular distributions for $\theta_c=30^{\circ};\,45^{\circ}$ and $60^{\circ}$ (from left to right).\\
\textit{Note:} The angular grid step for the PADs in (\textbf{d-i}) is similar to that in (\textbf{a-c}). The columns correspond to the photon energy indicated at the top of the figure: (\textbf{a,d,g}) -- $\omega=59.8$~eV; (\textbf{b,e,h}) -- $\omega=60.178$~eV; (\textbf{c,f,i}) -- $\omega=60.18$~eV. \label{fig2}}
\end{center}
\end{figure*}

Equation (\ref{eq:PADplane_generalS}) is the starting point for analyzing PADs generated in helium photo\-ionization by twisted radiation. Figures~\ref{fig2}a,b,c present calculated PADs with $P=0$ and $P=1$ for three different photon energies: 59.8~eV (below the resonance), 60.178 eV (just below the minimum in the $2s2p[^1P_1]$ dipole resonance), and 60.18 eV (exactly at the minimum).

In our further analysis, we follow the order of considerations in Section~\ref{Section3} and hence begin with circularly polarized twisted radiation. We assume $P=0$ in Eq.~(\ref{eq:PADplane_generalS}) and multiply each spherical harmonic $Y_{kq}(\theta_p,\phi_p)$ by the factor $d^k_{qq}(\theta_c)$ according to \textbf{Statement 1}. Although the statement refers to the Legendre polynomials,  for $P=0$ Eq.~(\ref{eq:PADplane_generalS}) contains only spherical harmonics $Y_{kq}(\theta_p,\phi_p)$ with $q=0$, which are equivalent to the Legendre polynomials. After such a transformation, we obtain the dependence of the PAD on the twisted radiation cone angle $\theta_c$. Simulated PADs for different values of $\theta_c$ are presented in Fig.~\ref{fig2}d,e,f. It is clearly seen that the PADs are very sensitive to both the photon energy and the angle~$\theta_c$. For $\omega=59.8$~eV in plane-wave photo\-ionization, the angular distribution is ``purely" dipole. Increasing $\theta_c$ leads to gains in the forward-backward direction, and the PAD becomes almost isotropic. Closer to the minimum of the dipole resonance ($\omega=60.178$~eV), the PADs start to lose the general symmetry because of amplification of non-dipole effects. Exactly in the minimum ($\omega=60.18$~eV according to our calculations), the shape of the PADs becomes qualitatively different. Specifically, a predominant portion of photo\-electrons is emitted in the direction of the incident beam wave vector $\bm{k}$ and a small fraction in the opposite direction.

Next we consider the linearly polarized case, set $P=1$ in Eq.~(\ref{eq:PADplane_generalS}) and multiply each spherical harmonic $Y_{kq}(\theta_p,\phi_p)$ by the factor $d^k_{qq}(\theta_c)$ according to \textbf{Statement 2}. Calculated PADs for this case are presented in Fig.~\ref{fig2}g,h,i. For $\omega=59.8$~eV, the evolutions of  the PADs with increasing $\theta_c$ do not show any striking changes, becoming only more intense in the forward-backward direction. On the contrary, a little below the dipole resonance minimum ($\omega=60.178$~eV), the angular distribution changes quite noticibly and a redistribution of photoelectrons  occurs. Finally, when the photon energy approaches the minimum at $\omega=60.18$~eV for the case of linearly polarized twisted light, we find significant changes in the shape of the PADs for different values of $\theta_c$. For $\theta_c=30^{\circ}$, for example, there are two dominant petals in the forward direction and two minor ones in the backward direction. Turning to $\theta_c=45^{\circ}$ merges the two backward petals into one and redistributes photo\-electrons in the forward direction by filling the local minimum along the incident beam wave vector $\bm{k}$. For $\theta_c=60^{\circ}$, two additional dominant backward directions of electron emission are formed ($\sim 67.5^{\circ}$ and $290.5^{\circ}$), and in the forward direction, the strengthening trend along the wave vector~$\bm{k}$ continues.

Summing up, the above analysis showed that the angular distributions of photo\-electrons emitted under the influence of twisted Bessel beam are very sensitive to the parameters of the incident radiation (polarization and cone angle $\theta_c$) in the energy regions where a strong domination of non-dipole effects occurs. Hence, one can control the shape of the PAD by manipulating the polarization and twisted radiation cone opening angle~$\theta_c$. 

From the above, it is clear that experimental angular distributions of high accuracy could serve as a tool to extract the parameters of twisted beams, i.e., one can diagnose the incident twisted radiation beam. Applying \textbf{Statement 1} to Eq.~(\ref{eq:PADplane_generalS}) for the case of circularly polarized ($P=0$) twisted beam and writing it in terms of Legendre polynomials, we obtain for the PAD:

\begin{widetext}
\begin{eqnarray} \label{eq:PADtw_circS}
\left( \frac{d\sigma^{(tw,circ)}}{d\Omega} \right)_{S} = \frac{\sigma_0}{4\pi} \left( 1-\frac{\beta}{2}P_{2}(\cos\theta_p)P_{2}(\cos\theta_c) + \left(\delta+\frac{\gamma}{5} \right)P_{1}(\cos\theta_p)P_{1}(\cos\theta_c) - \frac{\gamma}{5}P_{3}(\cos\theta_p)P_{3}(\cos\theta_c) 
\right).
\end{eqnarray}
\end{widetext}

\begin{figure}
\centering
\includegraphics[width=0.49\textwidth]{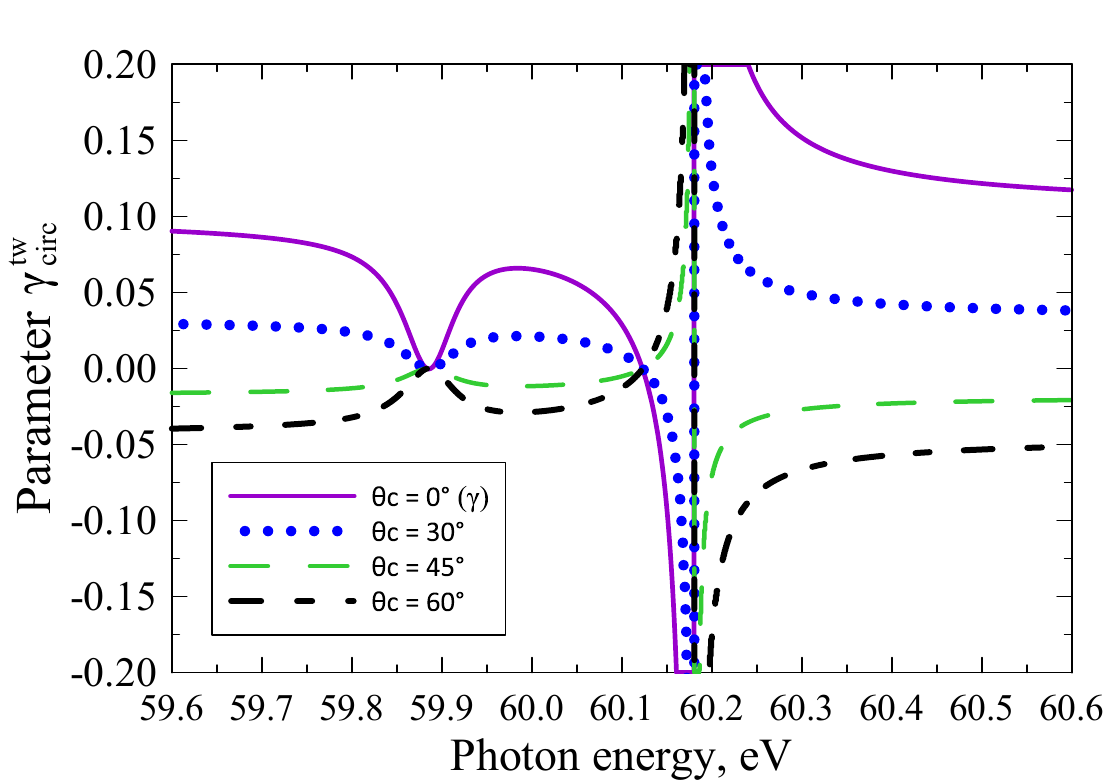}
\caption{Calculated parameterization factor $\gamma^{tw}_{circ}$ from Eq.~(\ref{eq:PADtw_circS_parametrized}) for different values of the twisted beam cone angle $\theta_c$ for helium photo\-ionization. For $\theta_c=0^{\circ}$, the factor $\gamma^{tw}_{circ}$ coincides with the conventional non-dipole parameter $\gamma$ depicted in Fig.~\ref{fig1}b.\label{fig5}}
\end{figure}

Putting
\begin{eqnarray}
\label{eq:parameterization1}
\beta^{tw}_{circ} &=& \beta P_{2}(\cos\theta_c), \\
\label{eq:parameterization2}
\gamma^{tw}_{circ} &=& \gamma P_{3}(\cos\theta_c), \\
\label{eq:parameterization3}
\delta^{tw}_{circ} &=& \left( \delta+\frac{\gamma}{5} \right)P_{1}(\cos\theta_c)-\frac{\gamma}{5}P_{3}(\cos\theta_c),
\end{eqnarray}
Eq.~(\ref{eq:PADtw_circS}) can be parameterized as

\begin{eqnarray} \label{eq:PADtw_circS_parametrized}
\left( \frac{d\sigma^{(tw,circ)}}{d\Omega} \right)_{S} = \frac{\sigma_0}{4\pi} \left( 1-\frac{\beta^{tw}_{circ}}{2}P_{2}(\cos\theta_p) + \right. \nonumber\\ + \left. \left(\delta^{tw}_{circ}+\frac{\gamma^{tw}_{circ}}{5} \right)P_{1}(\cos\theta_p) - \frac{\gamma^{tw}_{circ}}{5}P_{3}(\cos\theta_p) \right).
\end{eqnarray}
Equation~(\ref{eq:PADtw_circS_parametrized}) has the same structure as the PAD for photo\-ionization by  plane-wave circularly polarized radiation. Therefore, if one performs an experiment on photo\-ionization by both plane and twisted (Bessel) radiation with the same target atom and extracts the anisotropy parameters $\beta$, $\beta^{tw}_{circ}$, $\gamma$, $\gamma^{tw}_{circ}$, $\delta$, $\delta^{tw}_{circ}$, then it becomes possible to diagnose the Bessel beam, i.e., either to find  (according to parameterizations (\ref{eq:parameterization1})-(\ref{eq:parameterization3})) its parameter $\theta_c$, if it is unknown for some reason, or to estimate the quality of the twisted beam preparation by comparing the expected and the experimentally derived values of $\theta_c$. The dependence of $\gamma^{tw}_{circ}$ on the twisted cone angle $\theta_c$ is presented in Fig.~\ref{fig5}.

%%%%%%%%%%%%%%%%%%%%%%%%%%%%%%%%%%%%%%%%%%
%%%%%%%%%%%%%%%%%%%%%%%%%%%%%%%%%%%%%%%%%%
\section{Conclusions}

In the present work, we performed a theoretical analysis of the photo\-ionization process caused by twisted radiation, specifically Bessel beams. Assuming the atomic target to be extended (the size of the target area is larger than the characteristic size of the incident beam), we proved two statements that allow one to derive an expression for the PADs under the influence of twisted light of different polarization. Being extensions of the well-known parameterizations for the plane-wave radiation, our ``twisted” expressions will help to plan and to perform next-generation atomic photo\-ionization experiments. 

An illustration of the statements' application was given for the example of helium atoms ionized by twisted radiation in the vicinity of the lowest autoionization resonances in dipole and quadrupole photo\-ionization. When non-dipole effects become dominating, the shape of the PAD changes noticeably. Moreover, increasing the opening angle, i.e., the parameter $\theta_c$ for different incident photon energies, modifies the PADs substantially. The latter result suggests that the angular distributions can be controlled by the twisted radiation parameters. In addition, we showed that the PADs can serve as a diagnostic tool for the parameters of the incident circular polarized twisted Bessel radiation because of the possibility to parameterize the angular distribution expressions accordingly.

\begin{acknowledgments}
The authors benefited greatly from discussions with A.~Surzhykov and are acknowledge K.~Bartschat for careful reading of the manuscript and useful suggestions. The work on the development of formalism for the twisted light interaction with many-electron atoms and analysis of photo\-electron angular distributions in helium ionization by the Bessel light were funded by the Russian Science Foundation (project No. 21-42-04412; https://rscf.ru/en/project/21-42-04412/). The calculations of the photo\-ionization amplitudes were performed using resources of the Shared Services ``Data Center of the Far-Eastern Branch of the Russian Academy of Sciences” and supported by the Ministry of Science and Higher Education of the Russian Federation (project No. 0818-2020-0005).
\end{acknowledgments}

\appendix

\section{Differential cross section for circularly polarized Bessel light} \label{AppendixA}

By inserting the matrix element (\ref{eq:matrix_element_twisted}) into Eq.~(\ref{eq:crpss_section_circular}) we obtain:
\begin{widetext}
\begin{eqnarray}
    \label{eq:crpss_section_circular_final}
    \frac{{\rm d}\sigma^{\rm (tw, circ)}}{{\rm d}\Omega_p}\left(\theta_p, \phi_p ; \, \theta_c\right) &=&  \mathcal{N} \frac{1}{2J_i + 1} \times\, \nonumber \\[0.2cm]
    && \hspace*{-3.5cm} \times \sum\limits_{M_i M_f m_s}
    \int {\rm e}^{i({\bm k}'_\perp - {\bm k}_\perp) {\bm b}} \, 
    a_{\kappa_{\perp} m_{\mathsf{tam}}}({\bf k}_{\perp}) \, a^*_{\kappa_{\perp} m_{\mathsf{tam}}}({\bf k}'_{\perp}) \, M^{(\rm pl)}_{M_i \lambda M_f}\left({\bm k}, {\bm p} \right) \, M^{(\rm pl) *}_{M_i \lambda M_f}\left({\bm k}', {\bm p} \right) \,
    \frac{d^2 {\bf k}_{\perp}}{4 \pi^2} \, \frac{d^2 {\bf k}'_{\perp}}{4 \pi^2} \, 
    \frac{{\rm d}{\bm b}}{\pi R^2} = \nonumber \\[0.2cm]
    &=& \mathcal{N} \frac{1}{2J_i + 1} \,  \sum\limits_{M_i M_f m_s}
    \int \left| a_{\kappa_{\perp} m_{\mathsf{tam}}}({\bf k}_{\perp}) \right|^2 \,
    \left| M^{(\rm pl)}_{M_i \lambda M_f}\left({\bm k}, {\bm p} \right) \right|^2 \, \frac{d^2 {\bf k}_{\perp}}{4 \pi^2} \, .
\end{eqnarray}
\end{widetext}
By using the explicit form of the amplitude $a_{\kappa_{\perp} m_{\mathsf{tam}}}({\bf k}_{\perp})$ and the relation $\left|\delta(k_\perp - \kappa_\perp) \right|^2 = R/\pi \, \delta(k_\perp - \kappa_\perp)$ (cf.\ Eq.~(24) from Ref.~\cite{ScF14}), we finally obtain
\begin{eqnarray}
    \label{eq:crpss_section_circular_final_2}
 &&   \frac{{\rm d}\sigma^{\rm (tw, circ)}}{{\rm d}\Omega_p}\left(\theta_p, \phi_p ; \, \theta_c\right) = \nonumber\\
   & =&\mathcal{N} \frac{1}{2J_i + 1} \!\!\sum\limits_{M_i M_f m_s}
    \int \left| M^{(\rm pl)}_{M_i \lambda M_f}\left({\bm k}, {\bm p} \right) \right|^2  \frac{d {\varphi_k}}{2 \pi}.
\end{eqnarray}
Here, the plane-wave matrix element $M^{(\rm pl)}_{M_i \lambda M_f}\left({\bm k}, {\bm p} \right)$ is calculated for the photon wave--vector ${\bm k} = k \left(\sin\theta_c \cos\varphi_k, \, \sin\theta_c \sin\varphi_k, \, \cos\theta_c \right)$ with $k = c\omega$ and $\theta_c$ as input parameters.  
 
One can perform the integration over the azimuthal angle $\varphi_k$ analytically if one re\-writes the plane-wave matrix element (\ref{eq:matrix_element_plane_wave_final_2}) as
\begin{eqnarray}
    \label{eq:matrix_element_plane_wave_final_4}
M^{(\rm pl)}_{M_i \lambda M_f}\left({\bm k}, {\bm p} \right) =  \sum\limits_{LMp} \, \left(i \lambda \right)^{p} \, D^L_{M \lambda}(\hat{{\bm k}}) \, G_{L M}({\bm p}) \, , 
\end{eqnarray}
where we introduced 
\begin{eqnarray}
    \label{eq:matrix_element_plane_wave_final_G}
 &&    G_{L M}({\bm p}) = \sqrt{2\pi} \, i^L \, \sum\limits_{\kappa \mu} \, 
     \sum\limits_{J_t M_t}  \frac{[lL]}{[J_t]}\CGC{l}{0}{\frac{1}{2}}{m_s}{j}{m_s} \times \nonumber\\
   &  \times & D^{j *}_{\mu m_s}(\hat{\bm p}) \CGC{J_f}{M_f}{j}{\mu}{J_t}{M_t}\,  \CGC{J_i}{M_i}{L}{M}{J_t}{M_t}\times \nonumber\\
  &  \times &    \rmem{(\alpha_f J_f, \epsilon \kappa) J_t}{H_{\gamma}(pL)}{\alpha_i J_i} \, .
\end{eqnarray}
By inserting (\ref{eq:matrix_element_plane_wave_final_G}) into
(\ref{eq:matrix_element_plane_wave_final_4}) and using the relation 
$\int_0^\infty {\rm e}^{i(M-M')\varphi_k} {\rm d}\varphi_k = 2\pi \delta_{M M'}$, 
we obtain
\begin{eqnarray}
    \label{eq:crpss_section_circular_final_3}
&&    \frac{{\rm d}\sigma^{\rm (tw, circ)}}{{\rm d}\Omega_p}\left(\theta_p, \phi_p ; \, \theta_c\right) =  \mathcal{N} \frac{1}{2J_i + 1} \times \nonumber\\ 
  & \times & \sum\limits_{M_i M_f m_s} \sum\limits_{L L' p p'} \, \sum\limits_M  \left(i \lambda \right)^{p-p'} d^L_{M \lambda}(\theta_c) \, d^{L'}_{M \lambda}(\theta_c) \times \nonumber\\
  &  \times & G_{L M}({\bm p}) \, G^*_{L' M}({\bm p}) \, .
\end{eqnarray}
To find a more practical expression, we write  
Eq.~(\ref{eq:crpss_section_circular_final_3}) in the form
\begin{eqnarray} \label{eq:sectw1}
   && \frac{{\rm d}\sigma^{\rm (tw, circ)}}{{\rm d}\Omega_p}\left(\theta_p, \phi_p ; \, \theta_c\right) = \mathcal{N} \frac{2 \pi}{2J_i + 1} \times \nonumber\\
     &\times & \sum\limits_{L L' p p'} i^{L-L'} \left(i \lambda\right)^{p-p'} [LL'] \times \nonumber\\
     &\times &  \sum_{J_t J'_t \kappa \kappa'} {\cal Z} \, [ll'][J_tJ'_t]^{-1} \rmem{(\alpha_f J_f, \epsilon \kappa) J_t}{H_{\gamma}(pL)}{\alpha_i J_i} \times \nonumber\\
     &\times & \rmem{(\alpha_f J_f, \epsilon \kappa') J'_t}{H_{\gamma}(p'L')}{\alpha_i J_i}^{\ast},
\end{eqnarray}
where
\begin{eqnarray} \label{eq:z}
{\cal Z} &=& \sum\limits_{M_i M_f m_s M\atop \mu \mu' M_t M'_t}
    \CGC{l}{0}{\frac{1}{2}}{m_s}{j}{m_s}
    \CGC{J_f}{M_f}{j}{\mu}{J_t}{M_t}  \nonumber\\
  &\times&  \CGC{J_i}{M_i}{L}{M}{J_t}{M_t} \CGC{l'}{0}{\frac{1}{2}}{m_s}{j'}{m_s} 
    \CGC{J_f}{M_f}{j'}{\mu'}{J'_t}{M'_t}  \nonumber\\
 &\times&  \CGC{J_i}{M_i}{L'}{M}{J'_t}{M'_t} D^{j *}_{\mu m_s}(\hat{\bm p}) \, D^{j' }_{\mu' m_s}(\hat{\bm p}) \nonumber\\
 &\times&   d^L_{M \lambda}(\theta_c) \, d^{L'}_{M \lambda}(\theta_c) \,.
\end{eqnarray}
The summation over the projections in (\ref{eq:z}) can be performed analytically as follows: First we sum the product of four Clebsch-Gordan coefficients 
over $M_i M_f M_t M'_t$ using (A.91) of~\cite{Balashov2000}:
\begin{eqnarray} \label{eq:four}
&& \sum_{M_i M_f M_t M'_t}
    \CGC{J_f}{M_f}{j}{\mu}{J_t}{M_t} \, \CGC{J_i}{M_i}{L}{M}{J_t}{M_t}  \nonumber\\
  &\times& \CGC{J_f}{M_f}{j'}{\mu'}{J'_t}{M'_t} \, \CGC{J_i}{M_i}{L'}{M}{J'_t}{M'_t} = 
    \nonumber \\
 &=&\minus{J_t-J'_t+j-j'}  \sum_{s \sigma} [s J_t J'_t]^2 \, [jj']^{-1}
     \CGC{j}{-\!\!\mu}{s}{\sigma}{L}{-\!\!M}  \nonumber\\
 &\times&    \CGC{j'}{-\!\!\mu'}{s}{\sigma}{L'}{-\!\!M} \SechsJ{J_f}{J_t}{j}{L}{s}{J_i} \SechsJ{J_f}{J'_t}{j'}{L'}{s}{J_i}.
\end{eqnarray}
Then, multiplying $D$-functions, we obtain
\begin{eqnarray} \label{eq:prom1}
{\cal Z} & =&  \minus{J_t-J'_t+j-j'} [s J_t J'_t]^2 \, [jj']^{-1} \times \nonumber\\
&\times & \sum_{s m_s} \sum_{J M_1 M_2 M}  \sum_{\sigma \mu \mu'} \minus{m_s-\mu}
\CGC{j}{-\!\!\mu}{s}{\sigma}{L}{-\!\!M} \times \nonumber\\ 
& \times & \CGC{j'}{-\!\!\mu'}{s}{\sigma}{L'}{-\!\!M} \,
\CGC{j}{-\!\!\mu}{j'}{\mu'}{J}{M_1} \times \nonumber\\
&\times & \CGC{j}{-\!\!m_s}{j'}{m_s}{J}{M_2} \, D^{J}_{M_1 M_2}(\hat{\bm p}) \, \delta_{M_2 0} \CGC{l}{0}{\frac{1}{2}}{m_s}{j}{m_s} \times \nonumber\\
&\times & \CGC{l'}{0}{\frac{1}{2}}{m_s}{j'}{m_s} \, d^L_{M \lambda}(\theta_c) \, d^{L'}_{M \lambda}(\theta_c) \,.
\end{eqnarray}
Next, we sum the products of three Clebsch-Gordan coefficients over $\mu \mu' \sigma$:
\begin{eqnarray} \label{eq:3c1}
&& \sum_{\mu \mu' \sigma} \CGC{j}{-\!\!\mu}{s}{\sigma}{L}{-\!\!M}
\CGC{j'}{-\!\!\mu'}{s}{\sigma}{L'}{-\!\!M} \times \nonumber\\
&\times & \CGC{j}{-\!\!\mu}{j'}{\mu'}{J}{M_1} \minus{-\mu} = \minus{s-j+j'+J-M} \times \nonumber \\
&\times & [LL'] \CGC{L}{-\!\!M}{L'}{M}{J}{0}
\SechsJ{L}{L'}{J}{j'}{j}{s} \delta_{M_1 0} \, ,
\end{eqnarray}
and over $m_s$:
\begin{eqnarray} \label{eq:3c2}
& &\sum_{m_s} \minus{m_s} 
 \CGC{l}{0}{\frac{1}{2}}{m_s}{j}{m_s} \,
 \CGC{l'}{0}{\frac{1}{2}}{m_s}{j'}{m_s} \times \nonumber\\
& \times & \CGC{j}{-\!\!m_s}{j'}{m_s}{J}{0} = \minus{j'-j+J+\frac{1}{2}} [jj'] \times \nonumber\\
& \times & \CGC{l}{0}{l'}{0}{J}{0} \, \SechsJ{j}{j'}{J}{l'}{l}{\frac{1}{2}} \, .
\end{eqnarray}
The result is
\begin{eqnarray} \label{eq:3c3}
{\cal Z}& = &\sum_{sJM} \minus{J_t-J'_t++s+j'-j+J+\frac{1}{2}} \minus{J-M} [s J_t J'_t]^2 \times
\nonumber\\ &\times & [LL'] \, \CGC{l}{0}{l'}{0}{J}{0} \, \CGC{L}{-M}{L'}{M}{J}{0}
\nonumber\\
&&\SechsJ{J_f}{j}{J_t}{L}{J_i}{s} \SechsJ{J_f}{j'}{J'_t}{L'}{J_i}{s}
\SechsJ{j}{j'}{J}{L'}{L}{s} \SechsJ{j}{j'}{J}{l'}{l}{\frac{1}{2}} \nonumber\\
&&d^L_{M \lambda}(\theta_c) \, d^{L'}_{M \lambda}(\theta_c) \, P_J(\cos \theta_p)\,.
\end{eqnarray}
After this, we sum over $M$
\begin{eqnarray} \label{eq:3c4}
\sum_{M} \minus{J-M} \CGC{L}{-\!\!M}{L'}{M}{J}{0} \, d^L_{M \lambda}(\theta_c) \,
d^{L'}_{M \lambda}(\theta_c) = \nonumber\\
= \sum_M \minus{J-\lambda} 
\CGC{L}{-\!\!M}{L'}{M}{J}{0} \,
D^L_{-\!\!M -\!\!\lambda}(0, \theta_c,0) \times \nonumber\\ \times D^{L'}_{M \lambda}(0, \theta_c,0) = \minus{J-\lambda} \CGC{L}{-\!\!\lambda}{L'}{\lambda}{J}{0} \, P_J(\cos \theta_c) \, ,
\end{eqnarray}
and $s$:
\begin{eqnarray} \label{eq:3c5}
&&\sum_s \minus{s} [s]^2 \SechsJ{J_f}{j}{J_t}{L}{J_i}{s} \SechsJ{J_f}{j'}{J'_t}{L'}{J_i}{s}
\SechsJ{j}{j'}{J}{L'}{L}{s} \nonumber \\
&=& \minus{-J_f -J_i -L -j -j'-L'-J_t -J -J'_t} \times \nonumber \\
&&\SechsJ{J_t}{J}{J'_t}{j'}{J_f}{j} \SechsJ{J_t}{J}{J'_t}{L'}{J_i}{L} \, .
\end{eqnarray}
Collecting (\ref{eq:3c3})--(\ref{eq:3c5}), we arrive at
\begin{eqnarray} \label{eq:3c6}
{\cal Z}& =& \minus{J_i-J_f-\lambda-\frac{1}{2}} [J_t J'_t]^2 \, [LL'] \times \nonumber\\
&&\sum_J \minus{J} \CGC{l}{0}{l'}{0}{J}{0} \, \CGC{L}{\lambda}{L'}{-\!\!\lambda}{J}{0}
\nonumber\\
&&\SechsJ{J_t}{J'_t}{J}{j'}{j}{J_f} \SechsJ{J_t}{J'_t}{J}{L'}{L}{J_i} 
\SechsJ{j}{j'}{J}{l'}{l}{\frac{1}{2}} \nonumber\\&& P_J(\cos \theta_p) \, P_J(\cos \theta_c) \, .
\end{eqnarray}
Substituting~(\ref{eq:3c6}) into~(\ref{eq:sectw1}),
we finally obtain the cross section as
\begin{widetext}
\begin{eqnarray} %\label{eq:cr11}
 \frac{{\rm d}\sigma^{\rm (tw, circ)}}{{\rm d}\Omega_p}\left(\theta_p, \phi_p ; \, \theta_c\right)   & = & \mathcal{N} \frac{2 \pi}{2J_i + 1} 
 \sum_J \minus{J} P_J(\cos \theta_p) P_J(\cos \theta_c)
  \sum_{\kappa \kappa' J_t J'_t} \sum_{LL'pp'} i^{L-L'} [ll'J_tJ'_t] [LL']^2 \times \nonumber \\
 &&\times (i \lambda)^{p-p'}  
\minus{\lambda+J_i-J_f+\frac{1}{2}} \, 
 \CGC{L}{\lambda}{L'}{-\!\lambda}{J}{0} \, \CGC{l}{0}{l'}{0}{J}{0} 
 %\nonumber \\ 
  \SechsJ{j}{j'}{J}{l'}{l}{\frac{1}{2}} \SechsJ{J_t}{J'_t}{J}{L'}{L}{J_i}
  \SechsJ{J_t}{J'_t}{J}{j'}{j}{J_f} \times \nonumber\\  
 & & \times \rmem{(\alpha_f J_f, \epsilon \kappa) J_t}{H_{\gamma}(pL)}{\alpha_i J_i} \,
\rmem{(\alpha_f J_f, \epsilon \kappa') J'_t}{H_{\gamma}(p'L')}{\alpha_i J_i}^{\ast} \, .
\end{eqnarray}
\end{widetext}

\section{Differential cross section for linearly polarized Bessel light} \label{AppendixB}

Recall that $\hat{\bm p} \equiv \{ \phi_p, \theta_p, 0 \}$ and $\hat{\bm k} \equiv \{ \phi_k, \theta_c, 0 \}$. To simplify Eq.~(\ref{eq:crpss_section_linear}), we first substitute the matrix elements (\ref{eq:matrix_element_plane_wave_final_2}). The result is

\begin{widetext}
\begin{eqnarray} \label{eq:cr12}
&&\frac{{\rm d}\sigma^{\rm (tw, lin)}}{{\rm d}\Omega_p}\left(\theta_p, \phi_p ; \, \theta_c\right) = \frac{\mathcal{N}}{2J_i + 1} \, \sum\limits_{M_i M_f m_s}
\sum_{\lambda, \lambda'} \int \!d \phi_k \, e^{i(\lambda-\lambda') \phi_k}
\sum_{LL' J_t J'_t \atop pp' \kappa \kappa' } \sum_{MM' \mu \mu' \atop  M_t M'_t}
i^{L-L'}(i \lambda)^p (-i \lambda')^{p'} [ll'LL'] [J_t J'_t]^{-1} \times \nonumber \\
  &  \times & \CGC{l}{0}{\frac{1}{2}}{m_s}{j}{m_s} \, \CGC{l'}{0}{\frac{1}{2}}{m_s}{j'}{m_s} 
    \, \CGC{J_f}{M_f}{j}{\mu}{J_t}{M_t} \, \CGC{J_f}{M_f}{j'}{\mu'}{J'_t}{M'_t} \CGC{J_i}{M_i}{L}{M}{J_t}{M_t} \, \CGC{J_i}{M_i}{L'}{M'}{J'_t}{M'_t} \times \nonumber\\
    &\times & D^{j *}_{\mu m_s}(\hat{\bm p}) \, D^{j' }_{\mu' m_s}(\hat{\bm p}) \,
    D^{L}_{M \lambda}(\hat{\bm k}) \, D^{L' \ast}_{M' \lambda'}(\hat{\bm k}) \rmem{(\alpha_f J_f, \epsilon \kappa) J_t}{H_{\gamma}(pL)}{\alpha_i J_i} \,
\rmem{(\alpha_f J_f, \epsilon \kappa') J'_t}{H_{\gamma}(p'L')}{\alpha_i J_i}^{\ast}
\,.
\end{eqnarray}
\end{widetext}

Now we perform the summations
using the formulas
\begin{widetext}
\begin{eqnarray} \label{eq:four1}
&&\sum_{M_i M_f M_t M'_t}
   \CGC{J_f}{M_f}{j}{\mu}{J_t}{M_t} \, \CGC{J_i}{M_i}{L}{M}{J_t}{M_t}
    \CGC{J_f}{M_f}{j'}{\mu'}{J'_t}{M'_t} \, \CGC{J_i}{M_i}{L'}{M'}{J'_t}{M'_t} =
    \nonumber \\
   &  =& \minus{J_f-J_i-L+L'+j+j'-M'-\mu'} [J_t J'_t]^2 \sum_{s \sigma}
     \CGC{L'}{-\!\!M'}{L}{M}{s}{\sigma} \CGC{j'}{-\!\!\mu'}{j}{\mu}{s}{\sigma} \SechsJ{J_t}{J'_t}{s}{L'}{L}{J_i} \SechsJ{J_t}{J'_t}{s}{j'}{j}{J_f} \, ;
\end{eqnarray}

\begin{equation} \label{eq:dd1}
\sum_{\mu \mu'} \minus{-\mu'} \CGC{j'}{-\!\!\mu'}{j}{\mu}{s}{\sigma}
D^{j *}_{\mu m_s}(\hat{\bm p}) \, D^{j' }_{\mu' m_s}(\hat{\bm p}) 
= \minus{-m_s-2\sigma} \CGC{j}{-\!\!m_s}{j'}{m_s}{s}{0} 
D^{s}_{0 \sigma}(\hat{\bm p}^{-1}) \, ;
\end{equation}

\begin{equation} \label{eq:dd2}
\sum_{MM'} \minus{-M'} \CGC{L'}{-\!\!M'}{L}{M}{s}{\sigma}
D^{L}_{M \lambda}(\hat{\bm k}) \, D^{L' \ast}_{M' \lambda'}(\hat{\bm k}) 
= \minus{\lambda + \sigma} \CGC{L'}{-\!\!\lambda'}{L}{\lambda}{s}{\lambda-\lambda'} 
D^{s}_{\lambda-\lambda'  \sigma}(\hat{\bm k}^{-1}) \, .
\end{equation}
\end{widetext}

\noindent Here we introduced the notations $\hat{\bm p}^{-1} \equiv \{ 0, \theta_p, \phi_p \}$ and 
$\hat{\bm k}^{-1} \equiv \{ 0, \theta_c, \phi_k \}$.

\begin{eqnarray} \label{eq:three1}
&&\sum_{m_s} \minus{-m_s} \CGC{l}{0}{\frac{1}{2}}{m_s}{j}{m_s} \, \CGC{l'}{0}{\frac{1}{2}}{m_s}{j'}{m_s} \times \nonumber\\
&\times & \CGC{j}{-\!\!m_s}{j'}{m_s}{s}{0} = \minus{s-\frac{1}{2}} \CGC{l}{0}{l'}{0}{s}{0} 
\SechsJ{j}{j'}{s}{l'}{l}{\frac{1}{2}} \, . \nonumber\\
\end{eqnarray}

Finally, we use the integral
\begin{equation} \label{eq:igr}
\int \!d \phi_k \, e^{i(\lambda-\lambda') \phi_k} D^{s}_{\lambda-\lambda'  \sigma}(\hat{\bm k}^{-1}) = 2 \pi \, d^s_{\sigma \sigma}(\theta_c) 
\delta_{\sigma, \lambda-\lambda'}
\end{equation}
and note that
\begin{equation} \label{eq:dy}
D^{s}_{0 \sigma}(\hat{\bm p}^{-1}) = \minus{\sigma} \sqrt{\frac{4 \pi}{2s+1}}
Y^{\ast}_{s \sigma}(\hat{\bm p}) \, .
\end{equation}

Collecting (\ref{eq:cr12})-(\ref{eq:dy}) and replacing the summation indices
$s \sigma \rightarrow JM$, we obtain
\begin{widetext}
\begin{eqnarray}
 &&\frac{{\rm d}\sigma^{\rm (tw, lin)}}{{\rm d}\Omega_p}\left(\theta_p, \phi_p ; \, \theta_c\right) = \mathcal{N} \frac{2 \pi}{2J_i + 1} 
\sum_J \sum_{M=0, \pm2} \sqrt{\frac{4\pi}{2J+1}} Y^{\ast}_{JM}(\theta_p, \phi_p) \, d^J_{MM}(\theta_c) \sum_{\lambda, \lambda' } 
\delta_{M, \lambda-\lambda'} \times \nonumber\\
&\times & \sum_{\kappa \kappa' J_t J'_t}
\sum_{LL'pp'} i^{L-L'} (i \lambda)^p (-i \lambda')^{p'} \minus{J_i-J_f+\frac{1}{2}} \minus{L+\lambda} [ll'LL'J_t J'_t] \CGC{l}{0}{l'}{0}{J}{0} \CGC{L}{\lambda}{L'}{-\!\!\lambda'}{J}{M} \times \nonumber \\
&\times & \SechsJ{j}{j'}{J}{l'}{l}{\frac{1}{2}}
\SechsJ{J_t}{J'_t}{J}{L'}{L}{J_i} \SechsJ{J_t}{J'_t}{J}{j'}{j}{J_f}
 \rmem{(\alpha_f J_f, \epsilon \kappa) J_t}{H_{\gamma}(pL)}{\alpha_i J_i} \,
\rmem{(\alpha_f J_f, \epsilon \kappa') J'_t}{H_{\gamma}(p'L')}{\alpha_i J_i}^{\ast} \, .
\end{eqnarray}
\end{widetext}
%\nocite{*}

\bibliography{Twisted}% Produces the bibliography via BibTeX.

\providecommand{\noopsort}[1]{}\providecommand{\singleletter}[1]{#1}%
\begin{thebibliography}{43}%
\makeatletter
\providecommand \@ifxundefined [1]{%
 \@ifx{#1\undefined}
}%
\providecommand \@ifnum [1]{%
 \ifnum #1\expandafter \@firstoftwo
 \else \expandafter \@secondoftwo
 \fi
}%
\providecommand \@ifx [1]{%
 \ifx #1\expandafter \@firstoftwo
 \else \expandafter \@secondoftwo
 \fi
}%
\providecommand \natexlab [1]{#1}%
\providecommand \enquote  [1]{``#1''}%
\providecommand \bibnamefont  [1]{#1}%
\providecommand \bibfnamefont [1]{#1}%
\providecommand \citenamefont [1]{#1}%
\providecommand \href@noop [0]{\@secondoftwo}%
\providecommand \href [0]{\begingroup \@sanitize@url \@href}%
\providecommand \@href[1]{\@@startlink{#1}\@@href}%
\providecommand \@@href[1]{\endgroup#1\@@endlink}%
\providecommand \@sanitize@url [0]{\catcode `\\12\catcode `\$12\catcode
  `\&12\catcode `\#12\catcode `\^12\catcode `\_12\catcode `\%12\relax}%
\providecommand \@@startlink[1]{}%
\providecommand \@@endlink[0]{}%
\providecommand \url  [0]{\begingroup\@sanitize@url \@url }%
\providecommand \@url [1]{\endgroup\@href {#1}{\urlprefix }}%
\providecommand \urlprefix  [0]{URL }%
\providecommand \Eprint [0]{\href }%
\providecommand \doibase [0]{https://doi.org/}%
\providecommand \selectlanguage [0]{\@gobble}%
\providecommand \bibinfo  [0]{\@secondoftwo}%
\providecommand \bibfield  [0]{\@secondoftwo}%
\providecommand \translation [1]{[#1]}%
\providecommand \BibitemOpen [0]{}%
\providecommand \bibitemStop [0]{}%
\providecommand \bibitemNoStop [0]{.\EOS\space}%
\providecommand \EOS [0]{\spacefactor3000\relax}%
\providecommand \BibitemShut  [1]{\csname bibitem#1\endcsname}%
\let\auto@bib@innerbib\@empty
%</preamble>
\bibitem [{\citenamefont {Bahrdt}\ \emph {et~al.}(2013)\citenamefont {Bahrdt},
  \citenamefont {Holldack}, \citenamefont {Kuske}, \citenamefont {M\"uller},
  \citenamefont {Scheer},\ and\ \citenamefont {Schmid}}]{Bahrdt2013}%
  \BibitemOpen
  \bibfield  {author} {\bibinfo {author} {\bibfnamefont {J.}~\bibnamefont
  {Bahrdt}}, \bibinfo {author} {\bibfnamefont {K.}~\bibnamefont {Holldack}},
  \bibinfo {author} {\bibfnamefont {P.}~\bibnamefont {Kuske}}, \bibinfo
  {author} {\bibfnamefont {R.}~\bibnamefont {M\"uller}}, \bibinfo {author}
  {\bibfnamefont {M.}~\bibnamefont {Scheer}},\ and\ \bibinfo {author}
  {\bibfnamefont {P.}~\bibnamefont {Schmid}},\ }\bibfield  {title} {\bibinfo
  {title} {First observation of photons carrying orbital angular momentum in
  undulator radiation},\ }\href
  {https://doi.org/10.1103/PhysRevLett.111.034801} {\bibfield  {journal}
  {\bibinfo  {journal} {Phys. Rev. Lett.}\ }\textbf {\bibinfo {volume} {111}},\
  \bibinfo {pages} {034801} (\bibinfo {year} {2013})}\BibitemShut {NoStop}%
\bibitem [{\citenamefont {Molina-Terriza}\ \emph {et~al.}(2007)\citenamefont
  {Molina-Terriza}, \citenamefont {Torres},\ and\ \citenamefont
  {Torner}}]{Mollina-Terriza2007}%
  \BibitemOpen
  \bibfield  {author} {\bibinfo {author} {\bibfnamefont {G.}~\bibnamefont
  {Molina-Terriza}}, \bibinfo {author} {\bibfnamefont {J.~P.}\ \bibnamefont
  {Torres}},\ and\ \bibinfo {author} {\bibfnamefont {L.}~\bibnamefont
  {Torner}},\ }\bibfield  {title} {\bibinfo {title} {Twisted photons},\ }\href
  {https://doi.org/10.1038/nphys607} {\bibfield  {journal} {\bibinfo  {journal}
  {Nature Physics}\ }\textbf {\bibinfo {volume} {3}},\ \bibinfo {pages} {305}
  (\bibinfo {year} {2007})}\BibitemShut {NoStop}%
\bibitem [{\citenamefont {Bekshaev}\ \emph {et~al.}(2011)\citenamefont
  {Bekshaev}, \citenamefont {Bliokh},\ and\ \citenamefont
  {Soskin}}]{Bekshaev_2011}%
  \BibitemOpen
  \bibfield  {author} {\bibinfo {author} {\bibfnamefont {A.}~\bibnamefont
  {Bekshaev}}, \bibinfo {author} {\bibfnamefont {K.~Y.}\ \bibnamefont
  {Bliokh}},\ and\ \bibinfo {author} {\bibfnamefont {M.}~\bibnamefont
  {Soskin}},\ }\bibfield  {title} {\bibinfo {title} {Internal flows and energy
  circulation in light beams},\ }\href
  {https://doi.org/10.1088/2040-8978/13/5/053001} {\bibfield  {journal}
  {\bibinfo  {journal} {Journal of Optics}\ }\textbf {\bibinfo {volume} {13}},\
  \bibinfo {pages} {053001} (\bibinfo {year} {2011})}\BibitemShut {NoStop}%
\bibitem [{\citenamefont {Sueda}\ \emph {et~al.}(2004)\citenamefont {Sueda},
  \citenamefont {Miyaji}, \citenamefont {Miyanaga},\ and\ \citenamefont
  {Nakatsuka}}]{Sueda:04}%
  \BibitemOpen
  \bibfield  {author} {\bibinfo {author} {\bibfnamefont {K.}~\bibnamefont
  {Sueda}}, \bibinfo {author} {\bibfnamefont {G.}~\bibnamefont {Miyaji}},
  \bibinfo {author} {\bibfnamefont {N.}~\bibnamefont {Miyanaga}},\ and\
  \bibinfo {author} {\bibfnamefont {M.}~\bibnamefont {Nakatsuka}},\ }\bibfield
  {title} {\bibinfo {title} {Laguerre-gaussian beam generated with a multilevel
  spiral phase plate for high intensity laser pulses},\ }\href
  {https://doi.org/10.1364/OPEX.12.003548} {\bibfield  {journal} {\bibinfo
  {journal} {Opt. Express}\ }\textbf {\bibinfo {volume} {12}},\ \bibinfo
  {pages} {3548} (\bibinfo {year} {2004})}\BibitemShut {NoStop}%
\bibitem [{\citenamefont {Beijersbergen}\ \emph {et~al.}(1994)\citenamefont
  {Beijersbergen}, \citenamefont {Coerwinkel}, \citenamefont {Kristensen},\
  and\ \citenamefont {Woerdman}}]{BEIJERSBERGEN1994321}%
  \BibitemOpen
  \bibfield  {author} {\bibinfo {author} {\bibfnamefont {M.}~\bibnamefont
  {Beijersbergen}}, \bibinfo {author} {\bibfnamefont {R.}~\bibnamefont
  {Coerwinkel}}, \bibinfo {author} {\bibfnamefont {M.}~\bibnamefont
  {Kristensen}},\ and\ \bibinfo {author} {\bibfnamefont {J.}~\bibnamefont
  {Woerdman}},\ }\bibfield  {title} {\bibinfo {title} {Helical-wavefront laser
  beams produced with a spiral phaseplate},\ }\href
  {https://doi.org/https://doi.org/10.1016/0030-4018(94)90638-6} {\bibfield
  {journal} {\bibinfo  {journal} {Optics Communications}\ }\textbf {\bibinfo
  {volume} {112}},\ \bibinfo {pages} {321} (\bibinfo {year}
  {1994})}\BibitemShut {NoStop}%
\bibitem [{\citenamefont {Heckenberg}\ \emph {et~al.}(1992)\citenamefont
  {Heckenberg}, \citenamefont {McDuff}, \citenamefont {Smith},\ and\
  \citenamefont {White}}]{Heckenberg:92}%
  \BibitemOpen
  \bibfield  {author} {\bibinfo {author} {\bibfnamefont {N.~R.}\ \bibnamefont
  {Heckenberg}}, \bibinfo {author} {\bibfnamefont {R.}~\bibnamefont {McDuff}},
  \bibinfo {author} {\bibfnamefont {C.~P.}\ \bibnamefont {Smith}},\ and\
  \bibinfo {author} {\bibfnamefont {A.~G.}\ \bibnamefont {White}},\ }\bibfield
  {title} {\bibinfo {title} {Generation of optical phase singularities by
  computer-generated holograms},\ }\href {https://doi.org/10.1364/OL.17.000221}
  {\bibfield  {journal} {\bibinfo  {journal} {Opt. Lett.}\ }\textbf {\bibinfo
  {volume} {17}},\ \bibinfo {pages} {221} (\bibinfo {year} {1992})}\BibitemShut
  {NoStop}%
\bibitem [{\citenamefont {Karimi}\ \emph {et~al.}(2009)\citenamefont {Karimi},
  \citenamefont {Piccirillo}, \citenamefont {Nagali}, \citenamefont
  {Marrucci},\ and\ \citenamefont {Santamato}}]{Karimi2009}%
  \BibitemOpen
  \bibfield  {author} {\bibinfo {author} {\bibfnamefont {E.}~\bibnamefont
  {Karimi}}, \bibinfo {author} {\bibfnamefont {B.}~\bibnamefont {Piccirillo}},
  \bibinfo {author} {\bibfnamefont {E.}~\bibnamefont {Nagali}}, \bibinfo
  {author} {\bibfnamefont {L.}~\bibnamefont {Marrucci}},\ and\ \bibinfo
  {author} {\bibfnamefont {E.}~\bibnamefont {Santamato}},\ }\bibfield  {title}
  {\bibinfo {title} {Efficient generation and sorting of orbital angular
  momentum eigenmodes of light by thermally tuned q-plates},\ }\href
  {https://doi.org/10.1063/1.3154549} {\bibfield  {journal} {\bibinfo
  {journal} {Applied Physics Letters}\ }\textbf {\bibinfo {volume} {94}},\
  \bibinfo {pages} {231124} (\bibinfo {year} {2009})},\ \Eprint
  {https://arxiv.org/abs/https://doi.org/10.1063/1.3154549}
  {https://doi.org/10.1063/1.3154549} \BibitemShut {NoStop}%
\bibitem [{\citenamefont {Arlt}\ and\ \citenamefont
  {Dholakia}(2000)}]{ARLT2000297}%
  \BibitemOpen
  \bibfield  {author} {\bibinfo {author} {\bibfnamefont {J.}~\bibnamefont
  {Arlt}}\ and\ \bibinfo {author} {\bibfnamefont {K.}~\bibnamefont
  {Dholakia}},\ }\bibfield  {title} {\bibinfo {title} {Generation of high-order
  bessel beams by use of an axicon},\ }\href
  {https://doi.org/https://doi.org/10.1016/S0030-4018(00)00572-1} {\bibfield
  {journal} {\bibinfo  {journal} {Optics Communications}\ }\textbf {\bibinfo
  {volume} {177}},\ \bibinfo {pages} {297} (\bibinfo {year}
  {2000})}\BibitemShut {NoStop}%
\bibitem [{\citenamefont {Cai}\ \emph {et~al.}(2012)\citenamefont {Cai},
  \citenamefont {Wang}, \citenamefont {Strain}, \citenamefont {Johnson-Morris},
  \citenamefont {Zhu}, \citenamefont {Sorel}, \citenamefont {O’Brien},
  \citenamefont {Thompson},\ and\ \citenamefont {Yu}}]{Cai2012}%
  \BibitemOpen
  \bibfield  {author} {\bibinfo {author} {\bibfnamefont {X.}~\bibnamefont
  {Cai}}, \bibinfo {author} {\bibfnamefont {J.}~\bibnamefont {Wang}}, \bibinfo
  {author} {\bibfnamefont {M.~J.}\ \bibnamefont {Strain}}, \bibinfo {author}
  {\bibfnamefont {B.}~\bibnamefont {Johnson-Morris}}, \bibinfo {author}
  {\bibfnamefont {J.}~\bibnamefont {Zhu}}, \bibinfo {author} {\bibfnamefont
  {M.}~\bibnamefont {Sorel}}, \bibinfo {author} {\bibfnamefont {J.~L.}\
  \bibnamefont {O’Brien}}, \bibinfo {author} {\bibfnamefont {M.~G.}\
  \bibnamefont {Thompson}},\ and\ \bibinfo {author} {\bibfnamefont
  {S.}~\bibnamefont {Yu}},\ }\bibfield  {title} {\bibinfo {title} {Integrated
  compact optical vortex beam emitters},\ }\href
  {https://doi.org/10.1126/science.1226528} {\bibfield  {journal} {\bibinfo
  {journal} {Science}\ }\textbf {\bibinfo {volume} {338}},\ \bibinfo {pages}
  {363} (\bibinfo {year} {2012})},\ \Eprint
  {https://arxiv.org/abs/https://www.science.org/doi/pdf/10.1126/science.1226528}
  {https://www.science.org/doi/pdf/10.1126/science.1226528} \BibitemShut
  {NoStop}%
\bibitem [{\citenamefont {Yang}\ \emph {et~al.}(2021)\citenamefont {Yang},
  \citenamefont {Xie}, \citenamefont {He}, \citenamefont {Zhang},\ and\
  \citenamefont {Yuan}}]{Yang2021}%
  \BibitemOpen
  \bibfield  {author} {\bibinfo {author} {\bibfnamefont {H.}~\bibnamefont
  {Yang}}, \bibinfo {author} {\bibfnamefont {Z.}~\bibnamefont {Xie}}, \bibinfo
  {author} {\bibfnamefont {H.}~\bibnamefont {He}}, \bibinfo {author}
  {\bibfnamefont {Q.}~\bibnamefont {Zhang}},\ and\ \bibinfo {author}
  {\bibfnamefont {X.}~\bibnamefont {Yuan}},\ }\bibfield  {title} {\bibinfo
  {title} {A perspective on twisted light from on-chip devices},\ }\href
  {https://doi.org/10.1063/5.0060736} {\bibfield  {journal} {\bibinfo
  {journal} {APL Photonics}\ }\textbf {\bibinfo {volume} {6}},\ \bibinfo
  {pages} {110901} (\bibinfo {year} {2021})},\ \Eprint
  {https://arxiv.org/abs/https://doi.org/10.1063/5.0060736}
  {https://doi.org/10.1063/5.0060736} \BibitemShut {NoStop}%
\bibitem [{\citenamefont {Peele}\ \emph {et~al.}(2002)\citenamefont {Peele},
  \citenamefont {McMahon}, \citenamefont {Paterson}, \citenamefont {Tran},
  \citenamefont {Mancuso}, \citenamefont {Nugent}, \citenamefont {Hayes},
  \citenamefont {Harvey}, \citenamefont {Lai},\ and\ \citenamefont
  {McNulty}}]{Peele:02}%
  \BibitemOpen
  \bibfield  {author} {\bibinfo {author} {\bibfnamefont {A.~G.}\ \bibnamefont
  {Peele}}, \bibinfo {author} {\bibfnamefont {P.~J.}\ \bibnamefont {McMahon}},
  \bibinfo {author} {\bibfnamefont {D.}~\bibnamefont {Paterson}}, \bibinfo
  {author} {\bibfnamefont {C.~Q.}\ \bibnamefont {Tran}}, \bibinfo {author}
  {\bibfnamefont {A.~P.}\ \bibnamefont {Mancuso}}, \bibinfo {author}
  {\bibfnamefont {K.~A.}\ \bibnamefont {Nugent}}, \bibinfo {author}
  {\bibfnamefont {J.~P.}\ \bibnamefont {Hayes}}, \bibinfo {author}
  {\bibfnamefont {E.}~\bibnamefont {Harvey}}, \bibinfo {author} {\bibfnamefont
  {B.}~\bibnamefont {Lai}},\ and\ \bibinfo {author} {\bibfnamefont
  {I.}~\bibnamefont {McNulty}},\ }\bibfield  {title} {\bibinfo {title}
  {Observation of an x-ray vortex},\ }\href
  {https://doi.org/10.1364/OL.27.001752} {\bibfield  {journal} {\bibinfo
  {journal} {Opt. Lett.}\ }\textbf {\bibinfo {volume} {27}},\ \bibinfo {pages}
  {1752} (\bibinfo {year} {2002})}\BibitemShut {NoStop}%
\bibitem [{\citenamefont {Sasaki}\ and\ \citenamefont
  {McNulty}(2008)}]{Sasaki2008}%
  \BibitemOpen
  \bibfield  {author} {\bibinfo {author} {\bibfnamefont {S.}~\bibnamefont
  {Sasaki}}\ and\ \bibinfo {author} {\bibfnamefont {I.}~\bibnamefont
  {McNulty}},\ }\bibfield  {title} {\bibinfo {title} {Proposal for generating
  brilliant x-ray beams carrying orbital angular momentum},\ }\href
  {https://doi.org/10.1103/PhysRevLett.100.124801} {\bibfield  {journal}
  {\bibinfo  {journal} {Phys. Rev. Lett.}\ }\textbf {\bibinfo {volume} {100}},\
  \bibinfo {pages} {124801} (\bibinfo {year} {2008})}\BibitemShut {NoStop}%
\bibitem [{\citenamefont {Hemsing}\ \emph {et~al.}(2011)\citenamefont
  {Hemsing}, \citenamefont {Marinelli},\ and\ \citenamefont
  {Rosenzweig}}]{Hemsing2011}%
  \BibitemOpen
  \bibfield  {author} {\bibinfo {author} {\bibfnamefont {E.}~\bibnamefont
  {Hemsing}}, \bibinfo {author} {\bibfnamefont {A.}~\bibnamefont {Marinelli}},\
  and\ \bibinfo {author} {\bibfnamefont {J.~B.}\ \bibnamefont {Rosenzweig}},\
  }\bibfield  {title} {\bibinfo {title} {Generating optical orbital angular
  momentum in a high-gain free-electron laser at the first harmonic},\ }\href
  {https://doi.org/10.1103/PhysRevLett.106.164803} {\bibfield  {journal}
  {\bibinfo  {journal} {Phys. Rev. Lett.}\ }\textbf {\bibinfo {volume} {106}},\
  \bibinfo {pages} {164803} (\bibinfo {year} {2011})}\BibitemShut {NoStop}%
\bibitem [{\citenamefont {Jentschura}\ and\ \citenamefont
  {Serbo}(2011{\natexlab{a}})}]{Serbo2011}%
  \BibitemOpen
  \bibfield  {author} {\bibinfo {author} {\bibfnamefont {U.~D.}\ \bibnamefont
  {Jentschura}}\ and\ \bibinfo {author} {\bibfnamefont {V.~G.}\ \bibnamefont
  {Serbo}},\ }\bibfield  {title} {\bibinfo {title} {Compton upconversion
  of twisted photons: backscattering of particles with non-planar wave
  functions},\ }\href {https://doi.org/10.1140/epjc/s10052-011-1571-z}
  {\bibfield  {journal} {\bibinfo  {journal} {The European Physical Journal C}\
  }\textbf {\bibinfo {volume} {71}} (\bibinfo {year}
  {2011}{\natexlab{a}})}\BibitemShut {NoStop}%
\bibitem [{\citenamefont {Jentschura}\ and\ \citenamefont
  {Serbo}(2011{\natexlab{b}})}]{Serbo2011a}%
  \BibitemOpen
  \bibfield  {author} {\bibinfo {author} {\bibfnamefont {U.~D.}\ \bibnamefont
  {Jentschura}}\ and\ \bibinfo {author} {\bibfnamefont {V.~G.}\ \bibnamefont
  {Serbo}},\ }\bibfield  {title} {\bibinfo {title} {Generation of high-energy
  photons with large orbital angular momentum by compton backscattering},\
  }\href {https://doi.org/10.1103/PhysRevLett.106.013001} {\bibfield  {journal}
  {\bibinfo  {journal} {Phys. Rev. Lett.}\ }\textbf {\bibinfo {volume} {106}},\
  \bibinfo {pages} {013001} (\bibinfo {year} {2011}{\natexlab{b}})}\BibitemShut
  {NoStop}%
\bibitem [{\citenamefont {He}\ \emph {et~al.}(2013)\citenamefont {He},
  \citenamefont {Wang}, \citenamefont {Hu}, \citenamefont {Ye}, \citenamefont
  {Feng}, \citenamefont {Kan},\ and\ \citenamefont {Zhang}}]{He:13}%
  \BibitemOpen
  \bibfield  {author} {\bibinfo {author} {\bibfnamefont {J.}~\bibnamefont
  {He}}, \bibinfo {author} {\bibfnamefont {X.}~\bibnamefont {Wang}}, \bibinfo
  {author} {\bibfnamefont {D.}~\bibnamefont {Hu}}, \bibinfo {author}
  {\bibfnamefont {J.}~\bibnamefont {Ye}}, \bibinfo {author} {\bibfnamefont
  {S.}~\bibnamefont {Feng}}, \bibinfo {author} {\bibfnamefont {Q.}~\bibnamefont
  {Kan}},\ and\ \bibinfo {author} {\bibfnamefont {Y.}~\bibnamefont {Zhang}},\
  }\bibfield  {title} {\bibinfo {title} {Generation and evolution of the
  terahertz vortex beam},\ }\href {https://doi.org/10.1364/OE.21.020230}
  {\bibfield  {journal} {\bibinfo  {journal} {Opt. Express}\ }\textbf {\bibinfo
  {volume} {21}},\ \bibinfo {pages} {20230} (\bibinfo {year}
  {2013})}\BibitemShut {NoStop}%
\bibitem [{\citenamefont {Shen}\ \emph {et~al.}(2013)\citenamefont {Shen},
  \citenamefont {Campbell}, \citenamefont {Hage}, \citenamefont {Zou},
  \citenamefont {Buchler},\ and\ \citenamefont {Lam}}]{Shen_2013}%
  \BibitemOpen
  \bibfield  {author} {\bibinfo {author} {\bibfnamefont {Y.}~\bibnamefont
  {Shen}}, \bibinfo {author} {\bibfnamefont {G.~T.}\ \bibnamefont {Campbell}},
  \bibinfo {author} {\bibfnamefont {B.}~\bibnamefont {Hage}}, \bibinfo {author}
  {\bibfnamefont {H.}~\bibnamefont {Zou}}, \bibinfo {author} {\bibfnamefont
  {B.~C.}\ \bibnamefont {Buchler}},\ and\ \bibinfo {author} {\bibfnamefont
  {P.~K.}\ \bibnamefont {Lam}},\ }\bibfield  {title} {\bibinfo {title}
  {Generation and interferometric analysis of high charge optical vortices},\
  }\href {https://doi.org/10.1088/2040-8978/15/4/044005} {\bibfield  {journal}
  {\bibinfo  {journal} {Journal of Optics}\ }\textbf {\bibinfo {volume} {15}},\
  \bibinfo {pages} {044005} (\bibinfo {year} {2013})}\BibitemShut {NoStop}%
\bibitem [{\citenamefont {Ribi\ifmmode~\check{c}\else \v{c}\fi{}}\ \emph
  {et~al.}(2014)\citenamefont {Ribi\ifmmode~\check{c}\else \v{c}\fi{}},
  \citenamefont {Gauthier},\ and\ \citenamefont {De~Ninno}}]{Ribic2014}%
  \BibitemOpen
  \bibfield  {author} {\bibinfo {author} {\bibfnamefont {P.~c. v.~R.}\
  \bibnamefont {Ribi\ifmmode~\check{c}\else \v{c}\fi{}}}, \bibinfo {author}
  {\bibfnamefont {D.}~\bibnamefont {Gauthier}},\ and\ \bibinfo {author}
  {\bibfnamefont {G.}~\bibnamefont {De~Ninno}},\ }\bibfield  {title} {\bibinfo
  {title} {Generation of coherent extreme-ultraviolet radiation carrying
  orbital angular momentum},\ }\href
  {https://doi.org/10.1103/PhysRevLett.112.203602} {\bibfield  {journal}
  {\bibinfo  {journal} {Phys. Rev. Lett.}\ }\textbf {\bibinfo {volume} {112}},\
  \bibinfo {pages} {203602} (\bibinfo {year} {2014})}\BibitemShut {NoStop}%
\bibitem [{\citenamefont {Hernández-García}\ \emph
  {et~al.}(2017)\citenamefont {Hernández-García}, \citenamefont {Rego},
  \citenamefont {San~Román}, \citenamefont {Picón},\ and\ \citenamefont
  {Plaja}}]{hernandez2017}%
  \BibitemOpen
  \bibfield  {author} {\bibinfo {author} {\bibfnamefont {C.}~\bibnamefont
  {Hernández-García}}, \bibinfo {author} {\bibfnamefont {L.}~\bibnamefont
  {Rego}}, \bibinfo {author} {\bibfnamefont {J.}~\bibnamefont {San~Román}},
  \bibinfo {author} {\bibfnamefont {A.}~\bibnamefont {Picón}},\ and\ \bibinfo
  {author} {\bibfnamefont {L.}~\bibnamefont {Plaja}},\ }\bibfield  {title}
  {\bibinfo {title} {Attosecond twisted beams from high-order harmonic
  generation driven by optical vortices},\ }\href
  {https://doi.org/10.1017/hpl.2017.1} {\bibfield  {journal} {\bibinfo
  {journal} {High Power Laser Science and Engineering}\ }\textbf {\bibinfo
  {volume} {5}},\ \bibinfo {pages} {e3} (\bibinfo {year} {2017})}\BibitemShut
  {NoStop}%
\bibitem [{\citenamefont {Allen}\ \emph {et~al.}(1992)\citenamefont {Allen},
  \citenamefont {Beijersbergen}, \citenamefont {Spreeuw},\ and\ \citenamefont
  {Woerdman}}]{ALLEN1992}%
  \BibitemOpen
  \bibfield  {author} {\bibinfo {author} {\bibfnamefont {L.}~\bibnamefont
  {Allen}}, \bibinfo {author} {\bibfnamefont {M.~W.}\ \bibnamefont
  {Beijersbergen}}, \bibinfo {author} {\bibfnamefont {R.~J.~C.}\ \bibnamefont
  {Spreeuw}},\ and\ \bibinfo {author} {\bibfnamefont {J.~P.}\ \bibnamefont
  {Woerdman}},\ }\bibfield  {title} {\bibinfo {title} {Orbital angular momentum
  of light and the transformation of laguerre-gaussian laser modes},\ }\href
  {https://doi.org/10.1103/PhysRevA.45.8185} {\bibfield  {journal} {\bibinfo
  {journal} {Phys. Rev. A}\ }\textbf {\bibinfo {volume} {45}},\ \bibinfo
  {pages} {8185} (\bibinfo {year} {1992})}\BibitemShut {NoStop}%
\bibitem [{\citenamefont {Allen}\ \emph {et~al.}(1999)\citenamefont {Allen},
  \citenamefont {Padgett},\ and\ \citenamefont {Babiker}}]{ALLEN1999}%
  \BibitemOpen
  \bibfield  {author} {\bibinfo {author} {\bibfnamefont {L.}~\bibnamefont
  {Allen}}, \bibinfo {author} {\bibfnamefont {M.}~\bibnamefont {Padgett}},\
  and\ \bibinfo {author} {\bibfnamefont {M.}~\bibnamefont {Babiker}},\
  }\bibfield  {title} {\bibinfo {title} {Iv the orbital angular momentum of
  light}\ }(\bibinfo  {publisher} {Elsevier},\ \bibinfo {year} {1999})\ pp.\
  \bibinfo {pages} {291--372}\BibitemShut {NoStop}%
\bibitem [{\citenamefont {Durnin}(1987)}]{Durnin:87}%
  \BibitemOpen
  \bibfield  {author} {\bibinfo {author} {\bibfnamefont {J.}~\bibnamefont
  {Durnin}},\ }\bibfield  {title} {\bibinfo {title} {Exact solutions for
  nondiffracting beams. i. the scalar theory},\ }\href
  {https://doi.org/10.1364/JOSAA.4.000651} {\bibfield  {journal} {\bibinfo
  {journal} {J. Opt. Soc. Am. A}\ }\textbf {\bibinfo {volume} {4}},\ \bibinfo
  {pages} {651} (\bibinfo {year} {1987})}\BibitemShut {NoStop}%
\bibitem [{\citenamefont {Durnin}\ \emph {et~al.}(1987)\citenamefont {Durnin},
  \citenamefont {Miceli},\ and\ \citenamefont {Eberly}}]{DurninPRL}%
  \BibitemOpen
  \bibfield  {author} {\bibinfo {author} {\bibfnamefont {J.}~\bibnamefont
  {Durnin}}, \bibinfo {author} {\bibfnamefont {J.~J.}\ \bibnamefont {Miceli}},\
  and\ \bibinfo {author} {\bibfnamefont {J.~H.}\ \bibnamefont {Eberly}},\
  }\bibfield  {title} {\bibinfo {title} {Diffraction-free beams},\ }\href
  {https://doi.org/10.1103/PhysRevLett.58.1499} {\bibfield  {journal} {\bibinfo
   {journal} {Phys. Rev. Lett.}\ }\textbf {\bibinfo {volume} {58}},\ \bibinfo
  {pages} {1499} (\bibinfo {year} {1987})}\BibitemShut {NoStop}%
\bibitem [{\citenamefont {Babiker}\ \emph {et~al.}(2018)\citenamefont
  {Babiker}, \citenamefont {Andrews},\ and\ \citenamefont
  {Lembessis}}]{Babiker_2019}%
  \BibitemOpen
  \bibfield  {author} {\bibinfo {author} {\bibfnamefont {M.}~\bibnamefont
  {Babiker}}, \bibinfo {author} {\bibfnamefont {D.~L.}\ \bibnamefont
  {Andrews}},\ and\ \bibinfo {author} {\bibfnamefont {V.~E.}\ \bibnamefont
  {Lembessis}},\ }\bibfield  {title} {\bibinfo {title} {Atoms in complex
  twisted light},\ }\href {https://doi.org/10.1088/2040-8986/aaed14} {\bibfield
   {journal} {\bibinfo  {journal} {Journal of Optics}\ }\textbf {\bibinfo
  {volume} {21}},\ \bibinfo {pages} {013001} (\bibinfo {year}
  {2018})}\BibitemShut {NoStop}%
\bibitem [{\citenamefont {Surzhykov}\ \emph {et~al.}(2016)\citenamefont
  {Surzhykov}, \citenamefont {Seipt},\ and\ \citenamefont
  {Fritzsche}}]{Surzhykov2016}%
  \BibitemOpen
  \bibfield  {author} {\bibinfo {author} {\bibfnamefont {A.}~\bibnamefont
  {Surzhykov}}, \bibinfo {author} {\bibfnamefont {D.}~\bibnamefont {Seipt}},\
  and\ \bibinfo {author} {\bibfnamefont {S.}~\bibnamefont {Fritzsche}},\
  }\bibfield  {title} {\bibinfo {title} {Probing the energy flow in bessel
  light beams using atomic photoionization},\ }\href
  {https://doi.org/10.1103/PhysRevA.94.033420} {\bibfield  {journal} {\bibinfo
  {journal} {Phys. Rev. A}\ }\textbf {\bibinfo {volume} {94}},\ \bibinfo
  {pages} {033420} (\bibinfo {year} {2016})}\BibitemShut {NoStop}%
\bibitem [{\citenamefont {Kosheleva}\ \emph {et~al.}(2020)\citenamefont
  {Kosheleva}, \citenamefont {Zaytsev}, \citenamefont {M\"uller}, \citenamefont
  {Surzhykov},\ and\ \citenamefont {Fritzsche}}]{Kosheleva2020}%
  \BibitemOpen
  \bibfield  {author} {\bibinfo {author} {\bibfnamefont {V.~P.}\ \bibnamefont
  {Kosheleva}}, \bibinfo {author} {\bibfnamefont {V.~A.}\ \bibnamefont
  {Zaytsev}}, \bibinfo {author} {\bibfnamefont {R.~A.}\ \bibnamefont
  {M\"uller}}, \bibinfo {author} {\bibfnamefont {A.}~\bibnamefont
  {Surzhykov}},\ and\ \bibinfo {author} {\bibfnamefont {S.}~\bibnamefont
  {Fritzsche}},\ }\bibfield  {title} {\bibinfo {title} {Resonant two-photon
  ionization of atoms by twisted and plane-wave light},\ }\href
  {https://doi.org/10.1103/PhysRevA.102.063115} {\bibfield  {journal} {\bibinfo
   {journal} {Phys. Rev. A}\ }\textbf {\bibinfo {volume} {102}},\ \bibinfo
  {pages} {063115} (\bibinfo {year} {2020})}\BibitemShut {NoStop}%
\bibitem [{\citenamefont {Ramakrishna}\ \emph {et~al.}(2022)\citenamefont
  {Ramakrishna}, \citenamefont {Hofbrucker},\ and\ \citenamefont
  {Fritzsche}}]{Ramakrishna2022}%
  \BibitemOpen
  \bibfield  {author} {\bibinfo {author} {\bibfnamefont {S.}~\bibnamefont
  {Ramakrishna}}, \bibinfo {author} {\bibfnamefont {J.}~\bibnamefont
  {Hofbrucker}},\ and\ \bibinfo {author} {\bibfnamefont {S.}~\bibnamefont
  {Fritzsche}},\ }\bibfield  {title} {\bibinfo {title} {Photoexcitation of
  atoms by cylindrically polarized laguerre-gaussian beams},\ }\href
  {https://doi.org/10.1103/PhysRevA.105.033103} {\bibfield  {journal} {\bibinfo
   {journal} {Phys. Rev. A}\ }\textbf {\bibinfo {volume} {105}},\ \bibinfo
  {pages} {033103} (\bibinfo {year} {2022})}\BibitemShut {NoStop}%
\bibitem [{\citenamefont {Araoka}\ \emph {et~al.}(2005)\citenamefont {Araoka},
  \citenamefont {Verbiest}, \citenamefont {Clays},\ and\ \citenamefont
  {Persoons}}]{Araoka2005}%
  \BibitemOpen
  \bibfield  {author} {\bibinfo {author} {\bibfnamefont {F.}~\bibnamefont
  {Araoka}}, \bibinfo {author} {\bibfnamefont {T.}~\bibnamefont {Verbiest}},
  \bibinfo {author} {\bibfnamefont {K.}~\bibnamefont {Clays}},\ and\ \bibinfo
  {author} {\bibfnamefont {A.}~\bibnamefont {Persoons}},\ }\bibfield  {title}
  {\bibinfo {title} {Interactions of twisted light with chiral molecules: An
  experimental investigation},\ }\href
  {https://doi.org/10.1103/PhysRevA.71.055401} {\bibfield  {journal} {\bibinfo
  {journal} {Phys. Rev. A}\ }\textbf {\bibinfo {volume} {71}},\ \bibinfo
  {pages} {055401} (\bibinfo {year} {2005})}\BibitemShut {NoStop}%
\bibitem [{\citenamefont {Peshkov}\ \emph {et~al.}(2015)\citenamefont
  {Peshkov}, \citenamefont {Fritzsche},\ and\ \citenamefont
  {Surzhykov}}]{Peshkov2015}%
  \BibitemOpen
  \bibfield  {author} {\bibinfo {author} {\bibfnamefont {A.~A.}\ \bibnamefont
  {Peshkov}}, \bibinfo {author} {\bibfnamefont {S.}~\bibnamefont {Fritzsche}},\
  and\ \bibinfo {author} {\bibfnamefont {A.}~\bibnamefont {Surzhykov}},\
  }\bibfield  {title} {\bibinfo {title} {Ionization of ${\mathrm{h}}_{2}^{+}$
  molecular ions by twisted bessel light},\ }\href
  {https://doi.org/10.1103/PhysRevA.92.043415} {\bibfield  {journal} {\bibinfo
  {journal} {Phys. Rev. A}\ }\textbf {\bibinfo {volume} {92}},\ \bibinfo
  {pages} {043415} (\bibinfo {year} {2015})}\BibitemShut {NoStop}%
\bibitem [{\citenamefont {Matula}\ \emph {et~al.}(2013)\citenamefont {Matula},
  \citenamefont {Hayrapetyan}, \citenamefont {Serbo}, \citenamefont
  {Surzhykov},\ and\ \citenamefont {Fritzsche}}]{MaH13}%
  \BibitemOpen
  \bibfield  {author} {\bibinfo {author} {\bibfnamefont {O.}~\bibnamefont
  {Matula}}, \bibinfo {author} {\bibfnamefont {A.~G.}\ \bibnamefont
  {Hayrapetyan}}, \bibinfo {author} {\bibfnamefont {V.~G.}\ \bibnamefont
  {Serbo}}, \bibinfo {author} {\bibfnamefont {A.}~\bibnamefont {Surzhykov}},\
  and\ \bibinfo {author} {\bibfnamefont {S.}~\bibnamefont {Fritzsche}},\
  }\bibfield  {title} {\bibinfo {title} {Atomic ionization of hydrogen-like
  ions by twisted photons: {A}ngular distribution of emitted electrons},\
  }\href {https://doi.org/10.1088/0953-4075/46/20/205002} {\bibfield  {journal}
  {\bibinfo  {journal} {Journal of Physics B: Atomic, Molecular and Optical
  Physics}\ }\textbf {\bibinfo {volume} {46}},\ \bibinfo {pages} {205002}
  (\bibinfo {year} {2013})}\BibitemShut {NoStop}%
\bibitem [{\citenamefont {Seipt}\ \emph {et~al.}(2016)\citenamefont {Seipt},
  \citenamefont {M\"uller}, \citenamefont {Surzhykov},\ and\ \citenamefont
  {Fritzsche}}]{Seipt2016}%
  \BibitemOpen
  \bibfield  {author} {\bibinfo {author} {\bibfnamefont {D.}~\bibnamefont
  {Seipt}}, \bibinfo {author} {\bibfnamefont {R.~A.}\ \bibnamefont {M\"uller}},
  \bibinfo {author} {\bibfnamefont {A.}~\bibnamefont {Surzhykov}},\ and\
  \bibinfo {author} {\bibfnamefont {S.}~\bibnamefont {Fritzsche}},\ }\bibfield
  {title} {\bibinfo {title} {Two-color above-threshold ionization of atoms and
  ions in xuv bessel beams and intense laser light},\ }\href
  {https://doi.org/10.1103/PhysRevA.94.053420} {\bibfield  {journal} {\bibinfo
  {journal} {Phys. Rev. A}\ }\textbf {\bibinfo {volume} {94}},\ \bibinfo
  {pages} {053420} (\bibinfo {year} {2016})}\BibitemShut {NoStop}%
\bibitem [{\citenamefont {Pic\'{o}n}\ \emph {et~al.}(2010)\citenamefont
  {Pic\'{o}n}, \citenamefont {Mompart}, \citenamefont {de~Aldana},
  \citenamefont {Plaja}, \citenamefont {Calvo},\ and\ \citenamefont
  {Roso}}]{Picon:10}%
  \BibitemOpen
  \bibfield  {author} {\bibinfo {author} {\bibfnamefont {A.}~\bibnamefont
  {Pic\'{o}n}}, \bibinfo {author} {\bibfnamefont {J.}~\bibnamefont {Mompart}},
  \bibinfo {author} {\bibfnamefont {J.~R.~V.}\ \bibnamefont {de~Aldana}},
  \bibinfo {author} {\bibfnamefont {L.}~\bibnamefont {Plaja}}, \bibinfo
  {author} {\bibfnamefont {G.~F.}\ \bibnamefont {Calvo}},\ and\ \bibinfo
  {author} {\bibfnamefont {L.}~\bibnamefont {Roso}},\ }\bibfield  {title}
  {\bibinfo {title} {Photoionization with orbital angular momentum beams},\
  }\href {https://doi.org/10.1364/OE.18.003660} {\bibfield  {journal} {\bibinfo
   {journal} {Opt. Express}\ }\textbf {\bibinfo {volume} {18}},\ \bibinfo
  {pages} {3660} (\bibinfo {year} {2010})}\BibitemShut {NoStop}%
\bibitem [{\citenamefont {Cooper}(1990)}]{Cooper1990}%
  \BibitemOpen
  \bibfield  {author} {\bibinfo {author} {\bibfnamefont {J.~W.}\ \bibnamefont
  {Cooper}},\ }\bibfield  {title} {\bibinfo {title} {Multipole corrections to
  the angular distribution of photoelectrons at low energies},\ }\href
  {https://doi.org/10.1103/PhysRevA.42.6942} {\bibfield  {journal} {\bibinfo
  {journal} {Phys. Rev. A}\ }\textbf {\bibinfo {volume} {42}},\ \bibinfo
  {pages} {6942} (\bibinfo {year} {1990})}\BibitemShut {NoStop}%
\bibitem [{\citenamefont {Cooper}(1993)}]{Cooper1993}%
  \BibitemOpen
  \bibfield  {author} {\bibinfo {author} {\bibfnamefont {J.~W.}\ \bibnamefont
  {Cooper}},\ }\bibfield  {title} {\bibinfo {title}
  {Photoelectron-angular-distribution parameters for rare-gas subshells},\
  }\href {https://doi.org/10.1103/PhysRevA.47.1841} {\bibfield  {journal}
  {\bibinfo  {journal} {Phys. Rev. A}\ }\textbf {\bibinfo {volume} {47}},\
  \bibinfo {pages} {1841} (\bibinfo {year} {1993})}\BibitemShut {NoStop}%
\bibitem [{\citenamefont {Schulz}\ \emph {et~al.}(2019)\citenamefont {Schulz},
  \citenamefont {Fritzsche}, \citenamefont {M\"uller},\ and\ \citenamefont
  {Surzhykov}}]{Schulz2019}%
  \BibitemOpen
  \bibfield  {author} {\bibinfo {author} {\bibfnamefont {S.~A.-L.}\
  \bibnamefont {Schulz}}, \bibinfo {author} {\bibfnamefont {S.}~\bibnamefont
  {Fritzsche}}, \bibinfo {author} {\bibfnamefont {R.~A.}\ \bibnamefont
  {M\"uller}},\ and\ \bibinfo {author} {\bibfnamefont {A.}~\bibnamefont
  {Surzhykov}},\ }\bibfield  {title} {\bibinfo {title} {Modification of
  multipole transitions by twisted light},\ }\href
  {https://doi.org/10.1103/PhysRevA.100.043416} {\bibfield  {journal} {\bibinfo
   {journal} {Phys. Rev. A}\ }\textbf {\bibinfo {volume} {100}},\ \bibinfo
  {pages} {043416} (\bibinfo {year} {2019})}\BibitemShut {NoStop}%
\bibitem [{\citenamefont {Rose}(1957)}]{Ros57}%
  \BibitemOpen
  \bibfield  {author} {\bibinfo {author} {\bibfnamefont {M.~E.}\ \bibnamefont
  {Rose}},\ }\href@noop {} {\emph {\bibinfo {title} {Elementary theory of
  angular momentum}}}\ (\bibinfo  {publisher} {Wiley},\ \bibinfo {year}
  {1957})\BibitemShut {NoStop}%
\bibitem [{\citenamefont {Balashov}\ \emph {et~al.}(2000)\citenamefont
  {Balashov}, \citenamefont {Grum-Grzhimailo},\ and\ \citenamefont
  {Kabachnik}}]{Balashov2000}%
  \BibitemOpen
  \bibfield  {author} {\bibinfo {author} {\bibfnamefont {V.~V.}\ \bibnamefont
  {Balashov}}, \bibinfo {author} {\bibfnamefont {A.~N.}\ \bibnamefont
  {Grum-Grzhimailo}},\ and\ \bibinfo {author} {\bibfnamefont {N.~M.}\
  \bibnamefont {Kabachnik}},\ }\href
  {https://doi.org/10.1007/978-1-4757-3228-3} {\emph {\bibinfo {title}
  {Polarization and Correlation Phenomena in Atomic Collisions}}}\ (\bibinfo
  {publisher} {Springer US},\ \bibinfo {address} {Boston, MA},\ \bibinfo {year}
  {2000})\BibitemShut {NoStop}%
\bibitem [{\citenamefont {Scholz-Marggraf}\ \emph {et~al.}(2014)\citenamefont
  {Scholz-Marggraf}, \citenamefont {Fritzsche}, \citenamefont {Serbo},
  \citenamefont {Afanasev},\ and\ \citenamefont {Surzhykov}}]{ScF14}%
  \BibitemOpen
  \bibfield  {author} {\bibinfo {author} {\bibfnamefont {H.~M.}\ \bibnamefont
  {Scholz-Marggraf}}, \bibinfo {author} {\bibfnamefont {S.}~\bibnamefont
  {Fritzsche}}, \bibinfo {author} {\bibfnamefont {V.~G.}\ \bibnamefont
  {Serbo}}, \bibinfo {author} {\bibfnamefont {A.}~\bibnamefont {Afanasev}},\
  and\ \bibinfo {author} {\bibfnamefont {A.}~\bibnamefont {Surzhykov}},\
  }\bibfield  {title} {\bibinfo {title} {Absorption of twisted light by
  hydrogenlike atoms},\ }\href {https://doi.org/10.1103/PhysRevA.90.013425}
  {\bibfield  {journal} {\bibinfo  {journal} {Phys. Rev. A}\ }\textbf {\bibinfo
  {volume} {90}},\ \bibinfo {pages} {013425} (\bibinfo {year}
  {2014})}\BibitemShut {NoStop}%
\bibitem [{\citenamefont {Zatsarinny}(2006)}]{Zatsarinny2006}%
  \BibitemOpen
  \bibfield  {author} {\bibinfo {author} {\bibfnamefont {O.}~\bibnamefont
  {Zatsarinny}},\ }\href {https://doi.org/10.1016/j.cpc.2005.10.006} {\bibfield
   {journal} {\bibinfo  {journal} {Comp. Phys. Commun.}\ }\textbf {\bibinfo
  {volume} {174}},\ \bibinfo {pages} {273} (\bibinfo {year}
  {2006})}\BibitemShut {NoStop}%
\bibitem [{\citenamefont {Fischer}\ \emph {et~al.}(1997)\citenamefont
  {Fischer}, \citenamefont {Brage},\ and\ \citenamefont {Johnsson}}]{Froese97}%
  \BibitemOpen
  \bibfield  {author} {\bibinfo {author} {\bibfnamefont {C.~F.}\ \bibnamefont
  {Fischer}}, \bibinfo {author} {\bibfnamefont {T.}~\bibnamefont {Brage}},\
  and\ \bibinfo {author} {\bibfnamefont {P.}~\bibnamefont {Johnsson}},\
  }\href@noop {} {\emph {\bibinfo {title} {Computational Atomic Structure: An
  MCHF Approach}}}\ (\bibinfo  {publisher} {IOP Publishing: Bristol},\ \bibinfo
  {year} {1997})\BibitemShut {NoStop}%
\bibitem [{\citenamefont {Kr\"assig}\ \emph {et~al.}(2002)\citenamefont
  {Kr\"assig}, \citenamefont {Kanter}, \citenamefont {Southworth},
  \citenamefont {Guillemin}, \citenamefont {Hemmers}, \citenamefont {Lindle},
  \citenamefont {Wehlitz},\ and\ \citenamefont {Martin}}]{Krassig2002}%
  \BibitemOpen
  \bibfield  {author} {\bibinfo {author} {\bibfnamefont {B.}~\bibnamefont
  {Kr\"assig}}, \bibinfo {author} {\bibfnamefont {E.~P.}\ \bibnamefont
  {Kanter}}, \bibinfo {author} {\bibfnamefont {S.~H.}\ \bibnamefont
  {Southworth}}, \bibinfo {author} {\bibfnamefont {R.}~\bibnamefont
  {Guillemin}}, \bibinfo {author} {\bibfnamefont {O.}~\bibnamefont {Hemmers}},
  \bibinfo {author} {\bibfnamefont {D.~W.}\ \bibnamefont {Lindle}}, \bibinfo
  {author} {\bibfnamefont {R.}~\bibnamefont {Wehlitz}},\ and\ \bibinfo {author}
  {\bibfnamefont {N.~L.~S.}\ \bibnamefont {Martin}},\ }\bibfield  {title}
  {\bibinfo {title} {Photoexcitation of a dipole-forbidden resonance in
  helium},\ }\href {https://doi.org/10.1103/PhysRevLett.88.203002} {\bibfield
  {journal} {\bibinfo  {journal} {Phys. Rev. Lett.}\ }\textbf {\bibinfo
  {volume} {88}},\ \bibinfo {pages} {203002} (\bibinfo {year}
  {2002})}\BibitemShut {NoStop}%
\bibitem [{\citenamefont {Argenti}\ and\ \citenamefont
  {Moccia}(2010)}]{Argenti2010}%
  \BibitemOpen
  \bibfield  {author} {\bibinfo {author} {\bibfnamefont {L.}~\bibnamefont
  {Argenti}}\ and\ \bibinfo {author} {\bibfnamefont {R.}~\bibnamefont
  {Moccia}},\ }\bibfield  {title} {\bibinfo {title} {Nondipole effects in
  helium photoionization},\ }\href
  {https://doi.org/10.1088/0953-4075/43/23/235006} {\bibfield  {journal}
  {\bibinfo  {journal} {Journal of Physics B: Atomic, Molecular and Optical
  Physics}\ }\textbf {\bibinfo {volume} {43}},\ \bibinfo {pages} {235006}
  (\bibinfo {year} {2010})}\BibitemShut {NoStop}%
\bibitem [{\citenamefont {Shaw}\ \emph {et~al.}(1996)\citenamefont {Shaw},
  \citenamefont {Arp},\ and\ \citenamefont {Southworth}}]{Shaw1996}%
  \BibitemOpen
  \bibfield  {author} {\bibinfo {author} {\bibfnamefont {P.~S.}\ \bibnamefont
  {Shaw}}, \bibinfo {author} {\bibfnamefont {U.}~\bibnamefont {Arp}},\ and\
  \bibinfo {author} {\bibfnamefont {S.~H.}\ \bibnamefont {Southworth}},\
  }\bibfield  {title} {\bibinfo {title} {Measuring nondipolar asymmetries of
  photoelectron angular distributions},\ }\href
  {https://doi.org/10.1103/PhysRevA.54.1463} {\bibfield  {journal} {\bibinfo
  {journal} {Phys. Rev. A}\ }\textbf {\bibinfo {volume} {54}},\ \bibinfo
  {pages} {1463} (\bibinfo {year} {1996})}\BibitemShut {NoStop}%
\end{thebibliography}%

\end{document}